\documentclass[useAMS,usenatbib]{mnras}
\usepackage[dvips]{epsfig}
\usepackage[]{natbib}
\usepackage[]{fixltx2e}
\usepackage[]{amsmath}
\usepackage[]{amssymb}
\usepackage{enumerate}
\usepackage{array,multirow}
\usepackage{booktabs}
\usepackage{graphicx}

\title[Broad Line Region in local Seyferts]{Constraints on the Broad Line Region Properties and Extinction in Local Seyferts}

\author[A. Schnorr-M\"uller et al.]
  {Allan Schnorr-M\"uller$^{1,}$$^2$,
R.I.~Davies$^1$,
K.T.~Korista$^3$,
L.~Burtscher$^1$,
D.~Rosario$^4$,
\newauthor T.~Storchi-Bergmann$^5$,
A.~Contursi$^1$,
R.~Genzel$^1$,
J.~Graci\'a-Carpio$^1$,
E.K.S.~Hicks$^6$,
\newauthor A.~Janssen$^1$,
M.~Koss$^7$,
M.-Y.~Lin$^1$,
D.~Lutz$^1$,
W.~Maciejewski$^8$,
F.~M\"uller-S\'anchez$^9$,
\newauthor G.~Orban~de~Xivry$^1$,
R.~Riffel$^5$,
R.A.~Riffel$^{10}$,
M.~Schartmann$^{11}$,
A.~Sternberg$^{12}$,
\newauthor E.~Sturm$^{1}$,
L.~Tacconi$^{1}$,
S.~Veilleux$^{13}$,
O.~A.~Ulrich$^3$
\\
$^1$Max-Planck-Institut f\"ur extraterrestrische Physik, Giessenbachstr. 1, D-85741 Garching, Germany\\
$^2$CAPES Foundation, Ministry of Education of Brazil, 70040-020, Bras\'ilia, Brazil\\
$^3$Department of Physics, Western Michigan University, Kalamazoo, MI 49008-5252, USA\\
$^4$Centre of Extragalactic Astronomy, Department of Physics, Durham University, Durham, DH1 3LE, UK\\
$^5$Departamento de Astronomia, Universidade Federal do Rio Grande do Sul, IF, CP 15051, 91501-970 Porto Alegre, RS, Brazil\\
$^6$Physics \& Astronomy Department, University of Alaska, Anchorage, AK 99508-4664, USA\\
$^7$Institute for Astronomy, Department of Physics, ETH Zurich, Wolfgang-Pauli-Strasse 27, CH-8093 Zurich, Switzerland\\
$^8$Astrophysics Research Institute, Liverpool John Moores University, IC2 Liverpool Science Park, 146 Brownlow Hill, L3 5RF, UK\\
$^9$Center for Astrophysics and Space Astronomy, University of Colorado, Boulder, CO 80309-0389, USA\\
$^{10}$Departamento de F\'isica, Centro de Ci\^encias Naturais e Exatas, Universidade Federal de Santa Maria, 97105-900 Santa Maria, RS, Brazil\\
$^{11}$Centre for Astrophysics and Supercomputing, Swinburne University of Technology, Hawthorn, Victoria, 3122, Australia\\
$^{12}$Raymond and Beverly Sackler School of Physics \& Astronomy, Tel Aviv University, Ramat Aviv 69978, Israel\\
$^{13}$Department of Astronomy and Joint Space-Science Institute, University of Maryland, College Park, MD 20742-2421 USA
}

\begin{document}

\maketitle

\begin{abstract}
We use high spectral resolution (R\,$>8000$) data covering 3800--13000\r{A} to study the physical conditions of the broad line region (BLR) of nine nearby Seyfert 1 galaxies. 
Up to six broad H\,I lines are present in each spectrum. A comparison -- for the first time using simultaneous optical to near-infrared observations -- to photoionisation calculations with our devised simple scheme yields the extinction to the BLR at the same time as determining the density and photon flux, and hence distance from the nucleus, of the emitting gas.
This points to a typical density for the H\,I emitting gas of 10$^{11}$\,cm$^{-3}$ and shows that a significant amount of this gas lies at regions near the dust sublimation radius, consistent with theoretical predictions. 
We also confirm that in many objects the line ratios are far from case B, the best-fit intrinsic broad-line H$\alpha$/H$\beta$ ratios being in the range 2.5--6.6 as derived with our photoionization modeling scheme. The extinction to the BLR, based on independent estimates from H\,I and He\,II lines, is A$_V$\,$\le$\,3 for Seyfert 1--1.5s, while Seyfert 1.8--1.9s have A$_V$ in the range 4--8. A comparison of the extinction towards the BLR and narrow line region (NLR) indicates that the structure obscuring the BLR exists on scales smaller than the NLR. This could be the dusty torus, but dusty nuclear spirals or filaments could also be responsible.
The ratios between the X-ray absorbing column N$_H$ and the extinction to the BLR are consistent with the Galactic gas-to-dust ratio if N$_H$ variations are considered.
\end{abstract}

 
\section{Introduction}

The broad emission lines observed in the spectra of type 1 AGNs, including the partially obscured intermediate type AGN in which at least some broad lines are seen \citep{Osterbrock70}, are the most direct tracers of the activity and immediate environment of accreting supermassive black holes. 
However, despite the prominence of these lines, their use in deriving the physical conditions of the broad line region (BLR) proved troublesome until the development of the locally optimally emitting cloud (LOC) model by \cite{baldwin95}.
In this concept, the BLR comprises gas clouds covering a wide range of physical conditions; and line emission may arise from different regions of the ensemble according to where it is most optimally generated.
The gas `clouds' in the LOC model do not necessarily correspond to actual clouds, but may simply be packets of gas within a cloud that itself comprises gas with variety of properties.
As such, the more physically motivated models of the BLR, such as the radiation pressure confinement (RPC) model \citep{baskin14}, can also be considered LOC models.
The most general set of LOC models has been explored and compared to reverberation mapping results by \cite{korista00} and \cite{korista04}; and applied to quasars by \cite{korista12}.
Crucially, the generic models allow one to test the physical models such as RPC -- by indicating which part of the density vs photon-flux parameter space is consistent with the observed line emission, and comparing this to the specific prediction made by the physical model. Problems remain, however, in doing this because the intrinsic line ratios are not known {\it a priori} (the assumption of case B recombination as used for stellar photoionisation does not apply, primarily due to the high densities, high incident UV ionising fluxes, and optical depth of the emitted lines), and because of the intrinsic variability of the line emission on timescales of days to months.

If the BLR is totally unobscured, then the observed line ratios are the same as the intrinsic line ratios. In this context, \cite{lamura07} selected 90 Seyfert\,1 galaxies from the Sloan Digital Sky Survey (SDSS) in which all broad lines from H$\alpha$ to H$\delta$ could be measured -- since this implied negligible extinction to the BLR.
They derived $\langle$H$\alpha$/H$\beta\rangle = 3.45\pm0.65$, implying a wide range of intrinsic ratios, which they explored as a function of the line width.
With a similar aim, \cite{dong08} selected 446 Seyfert\,1s from the SDSS which had blue optical continuum, which implies low extinction to the accretion disk.
They found the distribution of H$\alpha$/H$\beta$ which could be well described by a log-Gaussian distribution peaking at 3.06 and with a standard deviation of only 0.03\,dex (corresponding to $\pm0.2$ in the ratio). This much tighter range of ratios led \cite{dong08} to suggest that one could adopt 3.06 as the intrinsic H$\alpha$/H$\beta$ and hence derive the extinction to the BLR. Recently, \citet{gaskell15} analysed a sample of low-redshift AGNs with extremely blue optical spectral indices and found that these objects have a mean  ratio of 2.72\,$\pm$\,0.04. They interpreted this result as an indication that case B recombination is valid in the BLR, and that even the bluest 10\% of SDSS AGNs still have significant reddening. On the other hand, \citet{baron16} analysed 5000 SDSS type 1 AGNs at z$\approx$\,0.4 and while they found that the majority of the objects in their sample are reddened, in unreddened objects the mean H$\alpha$/H$\beta$ ratio is $\approx$\,3 with a broad distribution in the range 1.5--4, which suggests that simple Case B recombination cannot explain the ratio in all sources. Despite these results, significant uncertainties remain. These include the impact on the observed H$\alpha$/H$\beta$ ratio of low continuum fluxes in unreddened AGN \citep{osterbrock81,rudy88,trippe08} and also of even low extinction to the BLR \citep{goodrich89,tram92,goodrich95,dong05,trippe10}.

Ideally one would like to derive the intrinsic line ratios and extinction simultaneously.
\cite{korista12} took a step in this direction, using observations of optical and near-infrared broad lines in 4 optically selected quasars -- although these were not obtained simultaneously.
These authors then compared the observed ratios to a grid of photoionisation models, in order to derive the best fitting density and photon flux.
By focussing on ratios involving near-infrared lines (Pa$\alpha$, Pa$\beta$, Pa$\gamma$) as well as H$\alpha$, the impact of extinction could be reduced.

In this paper we present high resolution spectra of Seyfert\,1s obtained simultaneously across the optical and infrared bands.
For the first time we are able to analyse ratios involving up to 6 broad H\,I lines (H$\gamma$, H$\beta$, H$\alpha$, Pa$\beta$, Pa$\gamma$ and Pa$\delta$) in order to derive both the intrinsic line ratios and the extinction for individual objects, by comparison to a grid of photoionisation calculations.
By using so many lines, we are also better able to account for the narrow line emission superimposed on the broad lines, in order to more robustly estimate the broad line fluxes.
And by focussing here on the hydrogen recombination lines, for which the physical process is in principle relatively simple, we avoid complications due to, for example, metallicity \citep{korista12}.
We note that additional problems associated with Ly$\alpha$ \citep{netzer95} are not relevant here since we do not make use of this line.

In Section\,\ref{Observations} we describe the observations and data reduction.
In Section\,\ref{sec:analysis} we describe the fitting of the broad and narrow emission lines, the photoionisation calculations and the extinction measurements.
Then in Section\,\ref{sec:discuss} we discuss the results, and in Section\,\ref{sec:conc} we present our conclusions.

\section {Observations and Data reduction}\label{Observations}

\subsection{Sample Selection}

The Seyfert 1 galaxies studied in this work are part of LLAMA (Local Luminous AGN with Matched Analogues), a study using VLT/SINFONI and VLT/Xshooter of a complete volume limited sample of nearby active galaxies selected by their 14--195\,keV
luminosity. The rationale for the selection of the LLAMA sample is described in detail in \citet{davies15}. 
Briefly, these authors note that the accretion rate, traced by the AGN luminosity, is a crucial parameter that is often ignored, especially in AGN samples selected via line ratios.
They argue that low luminosity AGN can in principle be powered by a very local gas reservoir on scales of $<10$\,pc; and it is only at higher luminosities that (observable) coherent dynamical mechanisms are required to drive sufficient gas inwards from circumnuclear scales of 100-1000\,pc.
If one wishes to study such mechanisms, the target galaxies must be near enough that these scales can be very well spatially resolved.
While there are many tracers that can be used to select local AGN based on luminosity, the very hard X-rays ($>10$\,keV) are widely accepted to be among the least biassed with respect to host galaxy properties.
The all-sky {\it Swift BAT} survey at 14-195\,keV therefore provides an excellent parent sample for the selection.
By selecting appropriate luminosity and distance (redshift) thresholds 
($\log{L_{14-195keV} [erg s^{-1}]} > 42.5$ and $z < 0.01$) one can select a complete volume limited sample from the flux limited survey.
An additional requirement that the objects are observable from the VLT lead to a sample of 20 AGN, of which 10 are classified as type 1.0-1.9 \citep{davies15}.

\subsection{Observations}

 \begin{table*}
  \begin{center}
    \begin{tabular}{c c c c c c c c c}
    \hline
    \hline
Object & Date(s) & Exposure & Air Mass & Seeing & AGN & log\,L$_{14-195}$ &log\,N$_{H}$ &Distance\\
      & Observed & (s) & & (\arcsec) & Classification & (erg\,s$^{-1}$) & (cm$^{-2}$) & (Mpc)\\
      \hline
      \multirow{2}{*}{MCG0514} & 23 Nov 2013 & 1920 & 1.012 & 0.61 & \multirow{2}{*}{Sy 1.0} & \multirow{2}{*}{42.60} & \multirow{2}{*}{$\le21.9^{1}$} & \multirow{2}{*}{41}\\
                & 11 Dec 2013 & 1920 & 1.002 & 0.72 & & & & \\
      MCG0523 & 22 Jan 2014 & 3840 & 1.063 & 1.21 & Sy 1.9 & 43.47 & 22.1-22.7$^2$ & 35\\
      MCG0630 & 16 Jun 2015 & 1920 & 1.124 & 0.83 & Sy 1.2 & 42.74 & 20.9$^1$-22.2$^3$ & 27\\
      NGC1365 & 10 Dec 2013 & 1920 & 1.022 & 1.34 & Sy 1.8 & 42.39 & 22.0-23.4$^{4}$ & 18\\
      \multirow{2}{*}{NGC2992} & 26 Feb 2014 & 3840 & 1.302 & 0.72 & \multirow{2}{*}{Sy 1.8} & \multirow{2}{*}{42.62} & \multirow{2}{*}{21.6-22.2$^2$} & \multirow{2}{*}{36}\\
       & 27 Feb 2014 & 3840 & 1.342 & 0.72 & & & &\\       
      NGC3783 & 11 Mar 2014 & 3840 & 1.428 & 0.81 & Sy 1.2 & 43.49 & 20.5$^1$-22.5$^5$ & 38\\
      NGC4235 & 13 May 2015 & 1920 & 1.203 & 0.73 & Sy 1.2 & 42.72 & 21.2$^6$-21.3$^1$ & 37\\
      NGC4593 & 10 Mar 2014 & 1920 & 1.325 & 0.80 & Sy 1.2 & 43.16 & 20.1$^7$-20.3$^5$ & 37\\
      NGC6814 & 13 May 2015 & 3840 & 1.077 & 0.86 & Sy 1.2 & 42.69 & 21.0$^1$-21.1$^3$ & 23\\
      \hline
    \end{tabular}
    \caption{Summary of the observations, Seyfert sub-types (see section\,3.4), 14-195\,kev luminosity (taken from \citealt{davies15}), X-ray absorbing column (for objects which show N$_{H}$ variations we show the smallest and largest values in the literature) and distance (taken from NED, using redshift independent estimates).
\textbf{Abbreviated names}:MCG--05-14-012, MCG--05-23-16 and MCG--06-30-015.
\textbf{References:}$^{1}$:\citet{davies15},$^{2}$:\citet{risaliti02},$^{3}$:\citet{molina09},$^{4}$:\citet{walton14},$^{5}$:\citet{lutz04},$^{6}$:\citet{papadakis08},$^{7}$:\citet{winter12}.}
   \label{table_obj}
  \end{center}
\end{table*}

High spectral resolutions were obtained using the X-shooter spectrograph mounted at the ESO-VLT in Paranal, Chile, between the nights of November 23, 2013 and June 16, 2015. A summary of the observations and basic properties of the sources are presented in Table~\ref{table_obj}. X-shooter is a multi wavelength high resolution spectrograph which consists of 3 spectroscopic arms, covering the 3000--5595\r{A} (UVB arm), 5595--10240\r{A} (VIS arm) and 10240--24800\r{A} (NIR arm) wavelength ranges. Each arm consists of an independent cross dispersed echelle spectrograph with its optimised optics, dispersive element and detector \citep{vernet11}. The observations were performed in IFU-offset mode, with a field-of-view of 1\farcs8$\,\times$\,4\arcsec. In IFU mode the resolving power of each arm is R\,$\approx$\,8400 for the UVB arm, R\,$\approx$\,13200 for the VIS arm and R\,$\approx$\,8300 for the NIR arm. The observations of each target consisted of two observing blocks with sky observations performed after each target observation. A spectroscopic standard star was observed on the same night as the target observation under similar atmospheric conditions. Telluric standard star observations were performed both before and after the target observations.

\subsection{Data Reduction}

The X-shooter spectra were reduced within the ESO Reflex environment \citep{freudling13}, using version 2.6.0 of the ESO X-shooter pipeline in IFU/Offset mode \citep{modigliani10}, in order to produce a 1\farcs8$\,\times$\,4\arcsec\ datacube. The reduction routine comprised detector bias and dark current subtraction, and subsequent rectification of the spectra and wavelength calibration. Relative and absolute flux calibration of the datacube were performed using data of a spectroscopic standard star observed on the same night as the source. We corrected the spectra for telluric absorption using a telluric standard star at a similar air mass and observed either before or right after the target observation.
A more detailed discussion of the specific issues encountered when processing these data, and of their solutions, will be given in Burtscher et al. (in prep).

\section{Data Analysis}\label{sec:analysis}

In this study, we make use of simultaneous observations of up to 6 broad and narrow H\,I lines at high spectral resolution ($R>8000$) to constrain physical properties of the broad line region and quantify the obscuration to the broad line and narrow line regions (BLR and NLR respectively) in a sample of nearby Seyfert~1s. Instead of using the whole datacube, we base our analysis on integrated spectra of the inner 1$\farcs2$, thus sampling only the innermost part of the NLR and reducing host galaxy contamination. The K band data is of lesser quality for all objects, thus Br$\gamma$ is not included in our analysis. Higher order Paschen and Balmer lines are present in the wavelength range covered by our observations, but these are mostly faint and only observed in a few objects, thus they are also not included in our analysis. 

Deriving the extinction from the broad H\,I line ratios cannot be done in a simple way, as photoionisation models predict that the line ratios may vary widely depending upon the conditions of the BLR \citep{netzer75,kwan81,korista04}.
Previous observational studies \citep{zhou06,lamura07,dong08} have measured the Balmer decrement in samples of type 1 AGNs, by selecting them based on the presence of a blue continuum \citep{dong08} or high order Balmer lines such as H$\delta$ and H$\epsilon$ \citep{lamura07}, in order to assure the extinction to the BLR is minimal. These studies found that the median H$\alpha$/H$\beta$ is 3.0--3.5, with moderate dispersions in the range 0.2--0.6 (mostly without correcting the dispersion caused by measurement errors). Nevertheless, ratios as small as 1.56 \citep{lamura07} and as large as 7.8 \citep{osterbrock81} have been observed in presumably unabsorbed AGNs. These results clearly show that case B recombination does not apply, and that there is a significant range of intrinsic H$\alpha$/H$\beta$ ratios.
For a meaningful measurement of the extinction to the BLR, one must allow for this range of ratios. To do so requires the use of photoionisation models covering a wide range of physical conditions, and observations of several H\,I line ratios. In this regard, the simultaneous coverage of up to 6 optical and near-infrared broad H\,I lines in our X-shooter observations provides an ideal opportunity for a detailed analysis of the extinction to, and the properties of, the BLR in Seyfert~1 galaxies at a level not possible in previous studies.

The broad He\,II$\lambda$4686\,\r{A} and He\,II\,$\lambda$10126.4\,\r{A} lines provide an alternative mean of assessing the extinction to the BLR where photoionisation models are not needed, as the ratio of these lines is expected to be similar to the case B prediction in the BLR spectrum \citep{bottorff02}. These broad He\,II lines are weak, however, and although He\,II\,$\lambda$10126.4\,\r{A} is observed in all but one object, broad He\,II\,$\lambda$4686\,\r{A} is observed in only four objects. Nevertheless, we derive the extinction to the BLR, or we estimate lower limits to the extinction if broad He\,II\,$\lambda$4686\,\r{A} is not observed, and compare these values to the ones derived from the H\,I lines. The O\,I\,$\lambda$11287/O\,I\,$\lambda$8446 provides another alternative for the determination of the extinction to the BLR \citep{rudy00}, however as O\,I\,$\lambda$11287\r{A} is only observed in a couple of sources, we do not make use of this line ratio in our analysis.     

In Sections~\ref{sec:emfit}--\ref{sec:extinction} we describe how we have measured the broad and narrow line fluxes from the data, outline the photoionisation calculations we use, and finally discuss the properties we can derive (extinction, photon flux, and gas density).

\subsection{Emission line fitting}
\label{sec:emfit}

In order to measure the broad and narrow Balmer line ratios, the narrow emission lines must be separated from the broad lines. Although this can be a relatively simple process in the case of Pa$\beta$, Pa$\gamma$ and H$\gamma$ (although \cite{landt14} discuss cases where this separation is not necessarily straightforward), it is not in the case of H$\alpha$ and H$\beta$. 
H$\alpha$ is observed blended with [N\,II] lines, which are usually not simple Gaussians but often asymmetric and frequently have faint extended wings at their bases. 
The broad H$\beta$ profile usually has superimposed on it an extended red wing which is not observed in other Balmer lines \citep{robertis85,marziani96}; it is believed to be formed, at least in part, by emission from weak Fe\,II multiplets and He\,I\,$\lambda$4922\r{A} \citep{kollatschny01,veron02}. For an accurate measurement of the line ratios, especially in the case of weak H$\alpha$ or H$\beta$, these features must be carefully taken into account. 
In Sections~\ref{sec:feii}--\ref{sec:fitbroad} we describe the procedures we have applied in order to deblend the broad lines from the other line profiles.

\subsubsection {FeII subtraction}
\label{sec:feii}

In the 4000--5600\r{A} wavelength range numerous Fe\,II multiplets are present, blending to form a pseudo-continuum and contaminating the continuum around H$\gamma$ and H$\beta$ and in some cases introducing an asymmetry in the H$\beta$ profile. Half of the sources presented significant Fe\,II emission, thus in these objects we first subtracted this emission before fitting the emission lines. To achieve this we performed a Levenberg-Marquardt least-squares minimisation using a theoretical template (see \citealt{kovacevic} for details) comprising five groups of Fe\,II lines. 
This was fitted to the observed spectra over spectral regions where no strong emission lines were present, namely 4150--4300\r{A}, 4450--4750\r{A} and 5100--5600\r{A}. The template was shifted and broadened (by convolution with Gaussians) and each of the five Fe\,II emission groups in the template was individually scaled. The result was subtracted from the observed spectra.

\subsubsection{Subtraction of broad He\,I\,$\lambda$4922\r{A}}
\label{sec:broadhei}

Broad He\,I\,$\lambda$4922\r{A} is observed blended with [O\,III]\,$\lambda$4959\r{A}, which makes it difficult to accurately determine its profile. Thus, instead of fitting Gaussians to the He\,I\,$\lambda$4922\r{A}, we fitted the broad He\,I\,$\lambda$5876 line with a combination of three Gaussians and adopted the resulting profile as a template for the He\,I\,$\lambda$4922\r{A} emission line, scaling the width of the template accordingly. We proceeded to fit the He\,I\,$\lambda$4922\r{A} and [O\,III]\,$\lambda$4959\r{A} blend with a combination of the He\,I template (where the flux scaling of the template was the only free parameter) and two Gaussians (which fitted the [O\,III]\,$\lambda$4959\r{A} profile).

\subsubsection{Narrow component subtraction}
\label{sec:narrow}

The narrow line subtraction was performed in two steps. In the first step, we performed a Levenberg-Marquardt least-squares minimisation where the narrow and broad H$\alpha$, H$\beta$, H$\gamma$ and P$\beta$ were fitted simultaneously. 
For the narrow lines, we created a template comprising two Gaussian components that was based on the [S\,II]\,$\lambda$6716\r{A} profile.
This template was used to fit the [N\,II] and narrow H\,I lines, by adjusting its flux scaling factor only.
The fits to the broad lines were subtracted from the spectra, isolating the narrow emission lines which were then re-fitted in the second step.   

In the second step, we fitted each narrow line with either one or two Gaussians, depending on whether there are inflections or broad wings in the [O\,III] profile (this way we are sure these features are real and not artefacts from poor broad line subtraction). We assumed the H\,I and [N\,II] lines have the same velocity dispersion and redshift and we set the flux of [N\,II]\,6583\,\r{A} line as 2.96 times that of the [N\,II]\,6548\r{A} line, in accordance with the ratio of
their transition probabilities \citep{osterbrock06}. The H$\alpha$ and [N\,II]\,6583\r{A} fluxes were free parameters in the fit.  We tied H$\gamma$, H$\beta$ and Pa$\beta$ fluxes to the H$\alpha$ flux together adopting fixed intrinsic line ratios and fitting for the extinction.
The extinction is likely to be different to that towards the BLR because the narrow lines originates in a physically distinct and larger region.
We estimate it from Balmer and Paschen hydrogen line ratios: H$\alpha$/H$\beta$, H$\gamma$/H$\beta$ and H$\alpha$/Pa$\beta$, adopting case B recombination line ratios assuming a temperature of 10000\,K and an electron density of 100\,cm$^{-3}$, except for the H$\alpha$/H$\beta$ ratio which we assumed is equal 3.1 due to collisional enhancement in the NLR.
We adopt a ``screen'' dust configuration and the \citet{wild11} extinction law, which  has the form: 
\begin{equation}
\frac{A_{\lambda}}{A_{V}}\,=\,0.6(\lambda/5500)^{-1.3} + 0.4(\lambda/5500)^{-0.7} 
\end{equation}
The first term describes extinction from a compact layer of dust, for which the exponent is set to match the extinction observed along lines of sight in the Milky-Way. The second term describes additional attenuation caused by the intervening diffuse interstellar medium (ISM). This law is a modification of the \citet{charlot00} law for star forming regions, and has been successfully adapted by \citet{wild11} so that it can be applied also to the narrow emission lines from AGNs. Through a comparison of unobscured mid-infrared  [Ne\,II]\,$\lambda$12.8$\mu$m\,+\,[Ne\,III]\,$\lambda$15.5$\mu$m emission lines to H$\alpha$, \citet{wild11} showed that this law provide a good correction to emission line fluxes even in active galaxies with a high dust content, motivating us to adopt this extinction law to estimate the extinction to the narrow and broad line regions in the present work. 

We point out that the observed narrow H\,I line ratios are not well fitted assuming case B recombination and adopting other extinction laws such as the \citet{gaskell07} law, the Small Magellanic Cloud law or the \citet{calzetti00} law. This occurs because A$_V$ values derived from the H$\alpha$/H$\beta$, H$\beta$/H$\gamma$ and H$\alpha$/Pa$\beta$ line ratios differ from one another when adopting either of these laws, although not by a large amount, while A$_V$ values derived from these ratios adopting the \citet{wild11} extinction law are consistent. Nevertheless, differences between the different extinction curves in the wavelength range covered in this study are not large (see Fig.\,2 in \citet{lyu14} for a comparison of these extinction curves).  

After completing the second fitting step above, we subtracted the fits to the narrow line from the spectra, isolating the broad lines. 

\subsubsection{Fitting the broad hydrogen lines}
\label{sec:fitbroad}

It is well known that broad line profiles are poorly represented by a single Gaussian. In order to create a more flexible profile, but limit the number of free parameters, we fitted the broad H$\alpha$, H$\beta$, H$\gamma$ and P$\beta$ lines simultaneously with three Gaussian components (except for NGC\,4593 where four Gaussian components were necessary). The redshift, velocity dispersion and relative flux of each Gaussian component was assumed to be the same for all emission lines, that is we assumed all broad H\,I lines have the same profile differing only by a scaling factor (except for NGC\,4593, where the broad H$\alpha$ line was fitted separately as its profile was significantly distinct from the other H\,I lines). 

We fitted Pa$\gamma$ and Pa$\delta$ separately from the other H\,I lines as both of them are strongly blended with other broad emission lines: He\,I\,$\lambda$10833.2\r{A} in the case of Pa$\gamma$, and He\,II\,$\lambda$10126.4\r{A} and Fe\,II\,$\lambda\lambda$9956,9998\r{A} in the case of Pa$\delta$. Before measuring the flux of the Pa$\gamma$ line, we subtracted the broad He\,I\,$\lambda$10833.2\r{A} from the spectra. This was done by adopting the fitted broad Pa$\beta$ profile as a template and scaling it so that it matches the blue wing of He\,I\,$\lambda$10833.2\r{A}. Subsequently, we fitted and scaled the broad Pa$\beta$ profile to the now unblended Pa$\gamma$ line. In the case of Pa$\delta$, we fitted and scaled the broad Pa$\beta$ profile to the peak of broad Pa$\delta$, subsequently subtracting the line from the spectra in order to isolate He\,II\,$\lambda$10126.4\r{A}. In the case of NGC\,3783 and NGC\,4593, we could not reliably separate Pa$\delta$ from He\,II\,$\lambda$10126.4\r{A} and Fe\,II\,$\lambda\lambda$9956,9998\r{A} using this technique so we excluded this line from our analysis of these objects. 

We show the resulting fits to all the broad lines in Figs.\,\ref{fig:mcg0523}--\ref{fig:ngc4593}.
Once the broad line fluxes had been measured, we compared their ratios to a grid of photoionisation models as described in Section~\ref{sec:photomodels}, in each case performing a minimisation to determine the best extinction as described in Section~\ref{sec:extinction}.

\subsubsection{Fitting the broad helium lines}
\label{sec:fitbroad_he}

The He\,II\,$\lambda$4686 and He\,II\,$\lambda$10126.4 lines were fitted simultaneously assuming they have the same profile. The narrow He\,II lines were fitted by a single Gaussian, while the broad lines were fitted by a combination of three Gaussians.

\subsection{Photoionization models}
\label{sec:photomodels}

We make use of the large grid of photoionisation models of constant cloud gas number density and solar gas abundances, similar to the grids presented in  \citet{korista04} and \citet{korista12}, computed with the photoionisation code Cloudy \citep{ferland13}, release c13.03. We refer the reader to the references above for a detailed description of the models, and give only a brief outline below.

Calculations of hydrogen emission line strengths were generated for a range of hydrogen number volume density ($n_{H}$) values of 10$^7$\,$\le$\,$n_{H}$\,$\le$\,10$^{14}$ and for a range of ionising photon flux ($\Phi_{H}$) values of 10$^{17}$\,$\le\,\Phi_{H}$\,$\le$\,10$^{24}$. The plane formed by these parameters represents the range of gas densities and distances from the ionising continuum that are expected to exist within, and account for the bulk of the line emission from, the BLR. For each point in the grid, Cloudy's standard AGN-ionising continuum was used. A constant number density within each cloud was assumed and a fixed column density of N$_{H}$\,=\,10$^{23}$\,cm$^{-2}$ was also adopted. Although a range of column densities is expected within the BLR, the H\,I emission line flux ratios are not strongly sensitive to the cloud column density in the range 10$^{22}$\,$<$\,N$_{H}$\,$<$10$^{24}$ \citep{korista97}.

The range of values for each of the physical parameters in our analysis is motivated by observations and physical considerations, following similar arguments as presented by \cite{korista12}. Regarding $n_{H}$, we adopt a lower limit of $n_{H}$\,$\ge$\,10$^{9}$\,cm$^{-3}$, based on the absence of broad forbidden lines and observations of broad C\,III] $\lambda$\,1909\r{A}, implying the presence of $\approx$\,10$^{9}$\,cm$^{-3}$ gas \citep{davidson79}. None of the broad emission lines gives an upper limit to the number density of the BLR, so we follow \citet{korista12} and adopt an upper limit of $n_{H}$\,=\,10$^{14}$\,cm$^{-3}$. 
We adopt a lower limit on $\Phi_{H}$ of 10$^{17}$\,cm$^{-2}$\,s$^{-1}$, which, for a fiducial AGN luminosity, is below that photon flux rate at the dust sublimation radius \citep{nenkova08}, beyond which the line emissivity is much lower \citep{netzer93}.
As the H\,I line emissivity is also low at large $\Phi_{H}$, we chose an upper limit on $\Phi_{H}$ of 10$^{24}$\,cm$^{-2}$\,s$^{-1}$. These limits yield a large range in the broad emission line region parameter space: 10$^9$\,$\le$\,$n_{H}$\,$\le$\,10$^{14}$ and 10$^{17}$\,$\le\,\Phi_{H}$\,$\le$\,10$^{24}$. Both parameters were stepped at intervals of 0.25\,dex.

In this work, our perspective is to directly compare the measured broad H\,I line ratios to the grid of photoionisation models described above, instead of applying LOC models to our data. An advantage of such approach is that we do not add model dependent constraints to our analysis, and thus we can use the data to constrain the range of valid grid points independent of any model for distributions of the properties of the clouds in the BLR. In fact, in the current analysis, our only constraint is to suppose that all the hydrogen lines are emitted from the same clouds with certain density and excitation conditions, which is a reasonable assumption considering the similarity of the profile of the various H\,I lines we analysed for each object. Additionally, if more than one grid point is a good match to the data, we adopt a simple weighting scheme where we calculate median $\Phi_{H}$, $n_{H}$ and line ratios from these point. A detailed description of our photoionization modelling scheme is given in Section\,\ref{sec:extinction}.

\subsection{Extinction Measurements}
\label{sec:extinction}

\begin{figure}

\includegraphics[scale=0.37,angle=-90]{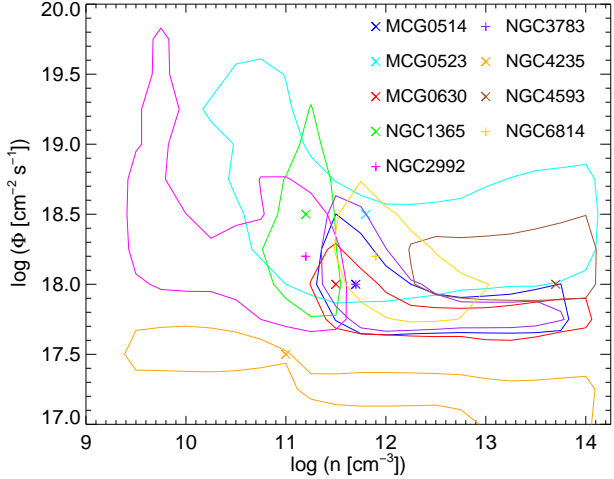}
\caption{Ionising photon flux $\Phi_{H}$ versus hydrogen number density $n_H$. The points show the location of best fit parameters for the objects in our sample. These are defined as the median of grid points where the difference between each model and de-reddened observed line ratios is less than 20$\%$. The extinction is optimised as a free parameter independently at each grid point, and the resulting values are listed in Table~\ref{tab:av}. The contours delimit the regions corresponding to 1$\sigma$ uncertainty on the derived parameters $n_H$ and $\Phi_H$. Details of the procedure are given in Section~\ref{sec:extinction}. Abbreviated name: MCG--05-14-012, MCG--05-23-16 and MCG--06-30-015.}
\label{fig1}
\end{figure}

\begin{figure*}

\includegraphics[scale=0.7,angle=-90]{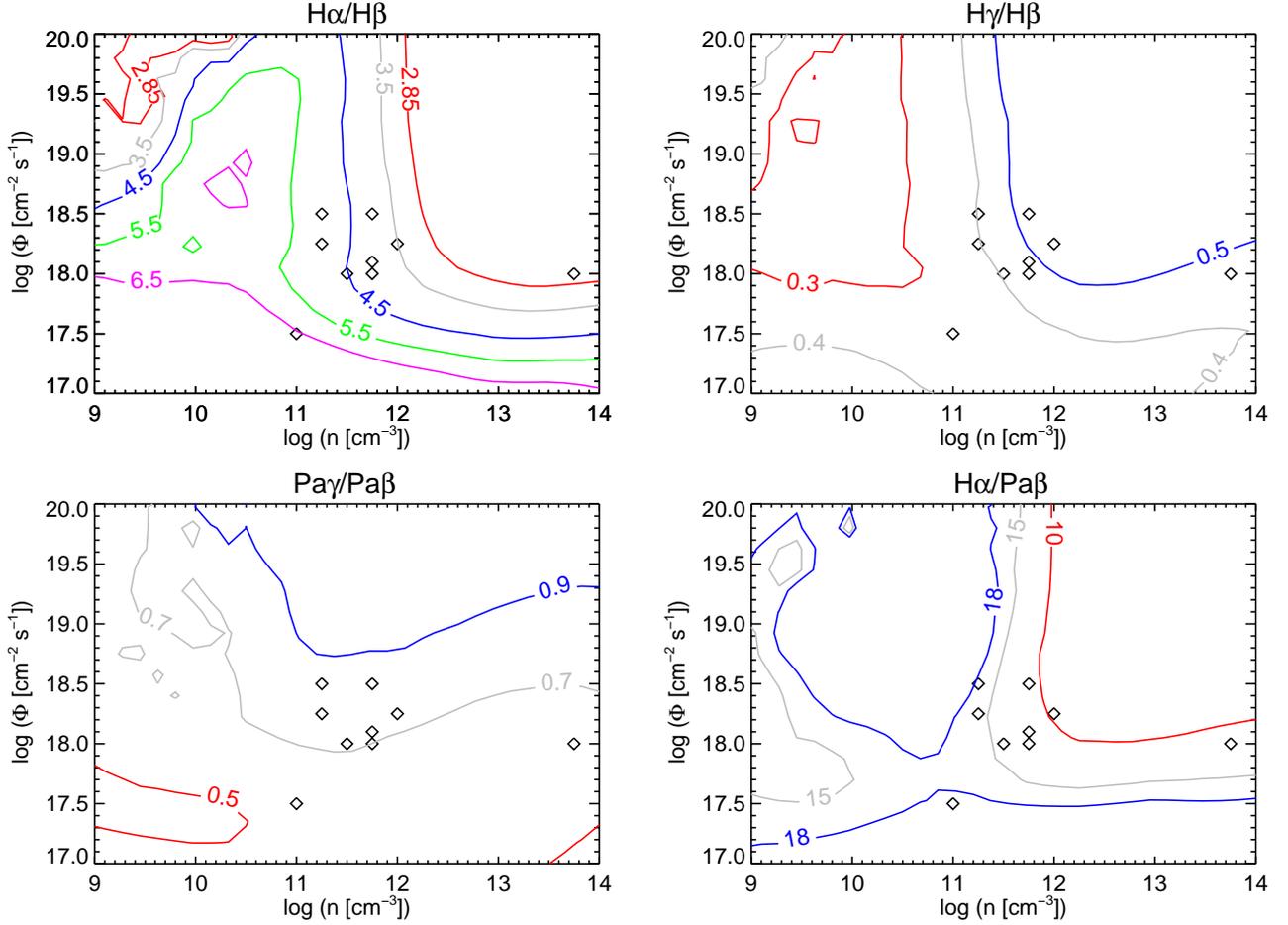}
\caption{Ionising photon flux $\Phi_{H}$ versus hydrogen number density $n_H$. The diamonds indicate the location of our AGN in the plane, as shown in Fig.~\ref{fig1}. 
Upper left panel: contours correspond to predicted H$\alpha$/H$\beta$ ratio values of 2.85, 3.5, 4.5, 5.5 and 6.5. Upper right panel: contours correspond to predicted H$\gamma$/H$\beta$ ratio values of 0.3, 0.5 and 0.7. Bottom left panel: contours correspond predicted to Pa$\gamma$/Pa$\beta$ ratio values of 0.5, 0.7 and 0.9. Bottom right panel: contours correspond to predicted H$\alpha$/Pa$\beta$ ratio values of 10.0, 15.0 and 18.0.}
\label{fig2}
\end{figure*}

\begin{table}
\centering
    \begin{tabular}{c c c c c  }
    \hline
    \hline
 Object&  \multicolumn{2}{c}{A$_{V}$(NLR)} & \multicolumn{2}{c}{A$_{V}$(BLR)} \\
 & H\,I & He\,II & H\,I & He\,II\\
      \hline
      \\
      MCG0514 & 0.7$^{+0.7}_{-0.7}$ &0.2$^{+0.3}_{-0.2}$  &  0.0$\pm0.2$ & 0.0$\pm0.3$\\
      \noalign{\vskip 2mm}
      MCG0523 & 1.6$^{+0.6}_{-0.8}$ & 1.3$^{+0.3}_{-0.4}$  & 7.7$\pm1.3$ & 4.2$^{*}$\\
      \noalign{\vskip 2mm}
      MCG0630 & 2.4$^{+0.7}_{-0.8}$ & 1.5$^{+0.3}_{-0.3}$ & 2.8$\pm0.4$ & 3.0$^{*}$\\
      \noalign{\vskip 2mm}
      NGC1365 & 1.3$^{+0.7}_{-0.7}$ & 2.2$^{+0.4}_{-0.3}$ & 4.2$\pm0.5$ & 4.4$^{*}$\\
      \noalign{\vskip 2mm}
      NGC2992 & 2.8$^{+0.7}_{-0.8}$ & 3.2$^{+0.3}_{-0.3}$ & 4.7$\pm0.5$ &  4.5$^{*}$\\
      \noalign{\vskip 2mm}
      NGC3783 & 0.0$^{+0.6}_{-0.0}$ &  - & 0.1$\pm0.2$ & -\\
      \noalign{\vskip 2mm}
      NGC4235 & 2.5$^{+0.7}_{-0.7}$ &  - & 1.5$\pm0.5$ & -\\
      \noalign{\vskip 2mm}
      NGC4593 & 0.0$^{+0.6}_{-0.0}$ & -  & 0.0$\pm0.1$ & -\\
      \noalign{\vskip 2mm}
      NGC6814 & 1.4$^{+0.6}_{-0.8}$ & 2.3$^{+0.4}_{-0.3}$ & 0.4$\pm0.4$ & 0.0$\pm0.3$\\
      \noalign{\vskip 2mm}
      \hline
    \end{tabular}
        \caption{Extinction to the narrow line region, A$_V$(NLR), and extinction to the broad line region, A$_V$(BLR), derived from Balmer and Paschen H\,I line ratios and the He\,II\,$\lambda$4686/He\,II\,$\lambda$10126.4 line ratio as described in Section~\ref{sec:extinction}. In objects where broad He\,II\,$\lambda$4686 is not detected a lower limit to A$_V$(BLR) was determined from broad He\,II\,$\lambda$10126.4\r{A}. These lower limits are marked with an asterisk. Abbreviated name: MCG--05-14-012, MCG--05-23-16 and MCG--06-30-015.}
    \label{tab:av}
\end{table}

\begin{table}
\centering
    \begin{tabular}{c c c }
    \hline
    \hline
Object & log($\Phi_{H}$) & log($n_H$) \\
      \hline
      MCG0514 & 18.0$\pm0.2$ & 11.7$\pm0.8$\\
      MCG0523 & 18.5$\pm0.5$ & 11.8$\pm1.3$\\
      MCG0630 & 18.0$\pm0.3$ & 11.5$\pm0.9$\\
      NGC1365 & 18.5$\pm0.7$ & 11.2$\pm0.9$\\
      NGC2992 & 18.2$\pm0.6$ & 11.2$\pm1.0$\\
      NGC3783 & 18.0$\pm0.3$ & 11.7$\pm0.9$\\
      NGC4235 & 17.5$\pm0.3$ & 11.0$\pm1.5$\\
      NGC4593 & 18.0$\pm0.2$ & 13.7$\pm0.6$\\      
      NGC6814 & 18.2$\pm0.8$ & 11.9$\pm1.3$\\
      \hline
    \end{tabular}
        \caption{Best fit parameters $\Phi_{H}$ and $n_H$.
Abbreviated name: MCG--05-14-012, MCG--05-23-16 and MCG--06-30-015.}
    \label{tab:phi}
\end{table}
                    
\begin{table*}
\centering
\tabcolsep=0.085cm
    \begin{tabular}{c c c c c c c c c c c c c c c c c c c c c c c}
    \hline
    \hline

\multirow{2}{*}{Object} & 
\multirow{2}{*}{A$_{V}$} & 
 \multicolumn{3}{c}{H$\gamma$/H$\beta$} & &
 \multicolumn{3}{c}{H$\alpha$/H$\beta$} &  & 
 \multicolumn{3}{c}{H$\alpha$/Pa$\beta$} & & 
 \multicolumn{3}{c}{Pa$\gamma$/Pa$\beta$} & &
 \multicolumn{3}{c}{Pa$\delta$/Pa$\beta$} & \\\cmidrule{3-5} \cmidrule{7-9} \cmidrule{11-13} \cmidrule{15-17} \cmidrule{19-21}
& (BLR)& m & u & p & & m & u & p & & m & u & p & & m & u & p & & m & u & p\\
      \hline
MCG0514 & 0.0$\pm0.2$&  0.40 &  -     & 0.46$\pm0.05$ & & 3.98 &    -  & 4.07$\pm0.76$ & & 12.7 & -     & 12.2$\pm1.93$ & & 0.57 & -    & 0.70$\pm0.07$ & & 0.50 & -    & 0.47$\pm0.07$\\
MCG0523 & 7.7$\pm1.3$&     - &  -     & 0.47$\pm0.13$ & &    - &    -  & 3.42$\pm1.61$ & & 0.74 & 13.5  & 13.9$\pm5.10$ & & 0.56 & 0.91 & 0.75$\pm0.09$ & & 0.24 & 0.56 & 0.54$\pm0.12$\\
MCG0630 & 2.8$\pm0.4$&  0.27 &  0.40  & 0.41$\pm0.04$ & & 8.85 & 3.98  & 4.41$\pm0.77$ & & 4.76 & 13.7  & 13.8$\pm2.60$ & & 0.65 & 0.77 & 0.63$\pm0.07$ & & 0.29 & 0.40 & 0.40$\pm0.07$\\
NGC1365 & 4.2$\pm0.5$&  0.24 &  0.43  & 0.40$\pm0.07$ & & 17.8 & 5.35  & 5.15$\pm0.74$ & & 3.34 & 16.3  & 16.4$\pm3.76$ & & 0.63 & 0.82 & 0.79$\pm0.13$ & & 0.40 & 0.63 & 0.57$\pm0.18$\\
NGC2992 & 4.7$\pm0.5$&      -&     -  & 0.39$\pm0.06$ & & 19.2 & 5.01  & 5.11$\pm1.00$ & & 2.71 & 16.0  & 15.8$\pm3.46$ & & 0.61 & 0.82 & 0.76$\pm0.08$ & & 0.29 & 0.48 & 0.53$\pm0.09$\\
NGC3783 & 0.1$\pm0.2$&  0.45 &     -  & 0.47$\pm0.06$ & & 4.10 & -     & 4.11$\pm0.77$ & & 12.9 & -     & 12.1$\pm2.10$ & & 0.72 & -    & 0.71$\pm0.09$ & & -    & -    & 0.48$\pm0.10$\\
NGC4235 & 1.5$\pm0.5$&  0.32 &  0.39  & 0.37$\pm0.04$ & & 10.2 & 6.66  & 6.58$\pm0.99$ & & 13.0 & 22.9  & 23.3$\pm4.42$ & &    - &    - & 0.55$\pm0.04$ & &    - &    - & 0.33$\pm0.03$\\
NGC4593 & 0.0$\pm0.1$&  0.49 &     -  & 0.49$\pm0.06$ & & 2.40 & -     & 2.49$\pm0.51$ & & 10.7 & -     & 10.5$\pm1.55$ & & 0.70 & -    & 0.64$\pm0.05$ & & -    & -    & 0.42$\pm0.05$\\
NGC6814 & 0.4$\pm0.4$&  0.50 &  0.58  & 0.49$\pm0.07$ & & 4.50 & 4.04  & 4.02$\pm0.77$ & & 9.56 & 11.0  & 11.0$\pm2.34$ & & 0.63 & 0.65 & 0.75$\pm0.09$ & & 0.53 & 0.55 & 0.53$\pm0.10$\\

      \hline
    \end{tabular}
        \caption{Best fit parameter A$_V$(BLR) and the measured (m), unreddened (u), and predicted (p) values for each line ratio. The `u' values are derived from the `m' values using the extinction given. A comparison of the `u' and `p' values for each ratio in each object indicates how well the observational measurements and photoionisation model predictions match.
Abbreviated name: MCG--05-14-012, MCG--05-23-16 and MCG--06-30-015.}
    \label{tab:ratio}
\end{table*}
      
The extinction affecting a set of line ratios may be determined by comparing the measured ratios with their theoretically predicted intrinsic values. 

In the case of the He\,II lines, considering the He\,II\,$\lambda$4686/He\,II\,$\lambda$10126.4 line ratio is expected to be similar to the case B prediction even in the BLR spectrum \citep{bottorff02}, we estimate the extinction to both the NLR and BLR assuming an intrinsic He\,II\,$\lambda$4686/He\,II\,$\lambda$10126.4 line ratio of 3.77, corresponding to the case B prediction for a temperature of 10000K and a density of 10$^6$\,cm$^{-3}$. 
If broad He\,II\,$\lambda$4686\r{A} is not observed, we estimate a lower limit to A$_V$(BLR) based on the broad He\,II\,$\lambda$10126.4\r{A} flux. This was done decreasing the observed broad He\,II\,$\lambda$4686/He\,II\,$\lambda$10126.4 ratio in steps of 0.1, starting from the intrinsic value of 3.77, until the broad He\,II\,$\lambda$4686\r{A} line becomes indistinguishable from the continuum. The lower limit to the extinction is then derived from this ratio. In Table~\ref{tab:av} we list A$_V$(NLR) and A$_V$(BLR) as estimated from the He\,II lines for each object. The errors in these estimates were derived assuming an uncertainty of 20$\%$ in the observed line ratios.

In the case of the H\,I lines, determining A$_V$(NLR), is a straightforward process since, as outlined in Section~\ref{sec:narrow}, a fixed set of intrinsic line ratios can be assumed. Deriving the extinction $A_V(NLR)$ to the NLR was done during the step of fitting and subtracting the narrow lines as described in that Section. In Table~\ref{tab:av} we list A$_V$(NLR) as estimated from the H\,I lines for each object. The errors in A$_V$ estimates were derived assuming an uncertainty of 20$\%$ in the observed line ratios.

The extinction to the BLR is derived in a different way, since the H\,I intrinsic line ratios are not known. To do so, we followed these steps:
\begin{enumerate}[1)]
\item We compared the measured emission line ratios to those derived from photoionisation calculations across the entire density-flux plane. At each individual point in the grid, we performed an independent Levenberg-Marquardt least-squares minimisation to derive the best fitting A$_{V}$ at that point. This results in a $\chi^2_{red}$ map showing how well the observations can be matched by photoionisation calculations for a wide range of density and incident flux.
\item As an estimate of the error in the broad line fluxes measurements is difficult to obtain, since it is dominated by systematic effects, we adopted an uncertainty of 20$\%$ for each line ratio (corresponding to a fiducial 10\% uncertainty on each line flux). From the resulting $\Delta\chi^{2}_{red}$ values, we are able to draw in the standard way the region corresponding to 1$\sigma$ uncertainty on the derived parameters $n_H$ and $\Phi_H$ as shown in Fig.~\ref{fig1}. 
\item In order to further constrain the range of valid grid points, we apply an additional constraint that the calculated ratios must all individually differ by less than 20$\%$ from the unreddened measured values. Grid points that did not fulfil this criterion were discarded. If more than one grid point fulfilled it, we calculated median values of $n_{H}$, $\Phi_{H}$, A$_V$(BLR) and line ratios. From here on we refer to the median values of the valid grid point as best fitting values. The best fitting values of $n_{H}$ and $\Phi_{H}$ are shown in Tab.~\ref{tab:phi}. The best fitting A$_V$(BLR) and predicted line ratios values are listed in Table~\ref{tab:ratio}. We note that the median values of $n_{H}$, $\Phi_{H}$ and A$_V$(BLR) are not significantly affected by increasing the selection threshold to 30$\%$. It is only when we allow the calculated and unreddened measured ratios to differ by 40$\%$ or more that the differences become significant, because at that point the density and photon flux become poorly constrained. We also note that the grid points fulfilling our chosen 20$\%$ criterion cover a much smaller range in the $n_{H}$-$\Phi_{H}$ plane than the range covered by the 1$\sigma$ uncertainty contours shown in Fig.~\ref{fig1}. Except for MCG--05-23-16 and NGC\,4235, where $n_{H}$ and $\Phi_{H}$ are poorly constrained because only three line ratios were measured for these objects, the number of valid grid points for each object is small, eight at most. Additionally the range of $n_{H}$ and $\Phi_{H}$ values fulfilling the criterion is small, the smallest and largest values differing by 1\,dex at most. 
\end{enumerate}

Considering that the extinctions to the NLR and BLR derived from the H\,I and He\,II line ratios are consistent, our analysis will focus on the A$_V$(NLR) and A$_V$(BLR) values derived from the HI lines.

In Fig.\,\ref{fig1} we plot the photon flux $\Phi_{H}$ versus hydrogen number density $n_H$ for each source, together with the 1$\sigma$ contours as described above.  In Fig.\,\ref{fig2} we show where these lie with respect to contours of the H$\alpha$/H$\beta$, H$\gamma$/H$\beta$, Pa$\gamma$/Pa$\beta$ and H$\alpha$/Pa$\beta$ line ratios in the $\Phi_{H}$-$n_H$ plane. The 1$\sigma$ uncertainty in the parameters $n_H$, $\Phi_H$, A$_V$(BLR) and line ratios were determined from the $\chi^2_{red}$ map (see Fig.\,\ref{fig1}). The best fitting values for photon flux $\Phi_{H}$, density $n_H$ and the respective  1$\sigma$ uncertainties are listed in Table~\ref{tab:phi}. In Table~\ref{tab:ratio} we list the best fit parameter A$_V$(BLR) and the measured (m), unreddened (u), and predicted (p) values for each line ratio. 
\subsection{Seyfert Sub-types}
\label{sec:subtypes}

Seyfert galaxies are typically divided into two categories, depending on whether their spectra show only narrow emission lines (Seyfert 2s), or also broad permitted emission lines (Seyfert 1s).
Some Seyferts, however, have characteristics intermediate between these two types, so additional categories were created based on the strength of the broad emission lines compared to the narrow emission lines \citep{osterbrock81}.
Various schemes have been proposed to classify these types, based on the ratio of the [O\,III]\,5007\r{A} flux to the total H$\beta$ flux. We adopt the scheme of \cite{whittle92} which has also been used by \citet{maiolino95}, noting that some of the thresholds differ slightly from those of \cite{winkler92} which were used by \cite{leonard15}.
The intermediate classifications are then:
\begin{itemize}
 \item Seyfert 1: Objects showing a broad H$\beta$ line with  [O\,III]/H$\beta$\,$<$\,0.3;
 \item Seyfert 1.2: Objects showing a broad H$\beta$ line with  0.3\,$<$\,[O\,III]/H$\beta$\,$<$\,1.0;
 \item Seyfert 1.5: Objects showing a broad H$\beta$ line with  1.0\,$<$\,[O\,III]/H$\beta$\,$<$\,4.0;
 \item Seyfert 1.8: Objects showing a broad H$\beta$ line with  4.0\,$<$\,[O\,III]/H$\beta$;
 \item Seyfert 1.9: Objects with no broad H$\beta$ line but showing a broad H$\alpha$ line.
\end{itemize}

We classify the objects in our sample as follows:
\begin{description}
\item{\bf MCG--05-23-16}: This object has been previously classified as a Seyfert 2 and as a near-infrared Seyfert 1 (where near-infrared, but not optical, broad lines are seen), but a careful subtraction of the narrow emission lines revealed a broad H$\alpha$ line. No broad H$\beta$ is observed, however. Thus we classify this object as a Seyfert 1.9.
\item{\bf MGC--05-14-012}: This object is a Seyfert 1 as [O\,III]/H$\beta$\,=\,0.2.
\item{\bf MGC--06-30-015}: This object is a Seyfert 1.2 as [O\,III]/H$\beta$\,=\,0.9 (but note that in the scheme of \cite{winkler92} this object would be a Seyfert 1.5).
\item{\bf NGC\,1365}: This object has previously been classified as a Seyfert 1.9 \citep{trippe10} and a Seyfert~1.8 \citep{schulz99}. A broad H$\beta$ line is present in our data and [O\,III]/H$\beta$\,=\,11.8, thus we classify it as a Seyfert 1.8.
\item{\bf NGC\,2992}: NGC\,2992 was classified as a Seyfert 1.9 by \citet{ward80} and more recently as a Seyfert~2 by \citet{trippe08}. An H$\alpha$ broad component is clearly present in our observations and after a careful subtraction of the narrow emission lines, broad H$\beta$ is also observed. [O\,III]/H$\beta$\,=\,10.4 in this object and thus we classify it as a Seyfert 1.8. 
\item{\bf NGC\,3783}: [O\,III]/H$\beta$\,=\,0.7 in this object and thus we classify it as a Seyfert~1.2 (noting that based on \cite{winkler92} this object would be a Seyfert 1.5).
\item{\bf NGC\,4235}: [O\,III]/H$\beta$\,=\,0.5 in this object, we classify it as a Seyfert 1.2 (noting that based on \cite{winkler92}, this is borderline between Seyfert 1.2 and 1.5).
\item{\bf NGC\,4593}: [O\,III]/H$\beta$\,=\,0.5 in this object, we classify it as a Seyfert 1.2 (noting that based on \cite{winkler92}, this is borderline between Seyfert 1.2 and 1.5).
\item{\bf NGC\,6814}: [O\,III]/H$\beta$\,=\,0.4 in this object, we classify it as a Seyfert 1.2.

\end{description}

\section{Discussion}
\label{sec:discuss}

In the previous Section, we identified the locations in the plane of photon flux versus density where, allowing for extinction, photoionisation models best match the observed broad line ratios. In doing so, we have derived $\Phi_H$, $n_H$, and $A_V(BLR)$ for each object.
In this section we explore the implications of that result.

\subsection{Photon flux, density, and location of the broad H\,I emitting region}
\label{sec:location}

The broad H\,I lines in our sample objects are, with the exception of NGC\,4593, all emitted in a region with densities in the range 10$^{11}$--10$^{12}$\,cm$^{-3}$.
For NGC\,4593 the preferred density is 10$^{13.7}$\,cm$^{-3}$.
However, the density is typically not strongly constrained and the 1$\sigma$ uncertainty, shown in Fig.~\ref{fig1}, encompasses densities that high for several objects.
\cite{korista12} also found high densities in the range 10$^{12}$--10$^{13.5}$\,cm$^{-3}$ for the four QSOs in which they analysed H\,I and He\,I line ratios.

However, if considered as a group, and taking into account the overlap between the 1$\sigma$ uncertainty ranges, more moderate densities of 10$^{11}$--10$^{12}$\,cm$^{-3}$ appear to typify our sample.
This is in remarkably good agreement with radiation pressure confinement (RPC) models \citep{baskin14} which predict a density of the order of $\approx$\,10$^{11}$\,cm$^{-3}$ near the hydrogen ionisation front.
The premise for these models is that if the gas pressure is determined by the radiation pressure then, in regions where strong H\,I emission occurs, the ionisation parameter will inevitable be $U \sim 0.1$ independent of the AGN luminosity and distance to the BLR.
Detailed modelling shows that in an irradiated slab of gas there will be a wide range of ionisation conditions, and thus high and low ionisation lines can be emitted from different locations within the same cloud.
Specifically, for clouds that are located half-way to the dust sublimation radius (and for an AGN luminosity of 10$^{45}$\,erg\,s$^{-1}$), the temperature falls to $\sim10^4$\,K at a column around 0.3--1$\times10^{23}$\,cm$^{-2}$ where the density is $\sim10^{11}$\,cm$^{-3}$. 
However, for clouds much closer to the same AGN, the density is much higher, $\sim10^{14}$\,cm$^{-3}$ where this temperature is reached and so collisional effects significantly reduce the resulting H\,I line strength.

The range of densities found for the H\,I emitting gas in this work is larger, however, than the upper limit derived from observations of C\,III]\,$\lambda$\,1909\r{A}, which is collisionally suppressed at densities larger than 10$^{9.5}$\,cm$^{-3}$. Reverberation mapping studies of nearby AGN found that both H$\beta$ and C\,III]\,$\lambda$\,1909\r{A} have similar time lags, implying they are produced at similar radii, in the outer part of the BLR \citep{onken02,trevese14}, thus one might expect the H\,I emitting gas and C\,III]\,$\lambda$\,1909\r{A} have similar densities. 
Regarding this issue, we note that a broad range of densities is expected in a single cloud with N$_{H}$\,$\approx$\,10$^{23}$\,cm$^{-2}$, so a density of 10$^{11}$\,cm$^{-3}$ for H$\beta$ and 10$^{9.5}$\,cm$^{-3}$ for C\,III]\,$\lambda$\,1909\r{A} is consistent with reverberation mapping results.
For an additional discussion on broad C\,III]\,$\lambda$\,1909\r{A} and H$\beta$, we refer to \citet{korista97} where maps of emission line equivalent widths in the density--ionising photon flux plane are shown. There one can see that the gas that is responsible for C\,III]\,$\lambda$\,1909\r{A} contributes to H$\beta$ at a typical fraction of $\approx$\,20--40\% of that found in gas at densities $>$\,10$^{11}$\,cm$^{-3}$, all else being equal (see their Figs. 3e and 3g). Additionally, it is important to consider that H$\beta$ is emitted in most gas clouds that emit C\,III]\,$\lambda$\,1909\r{A}, but the reverse is not true. So in this sense, the present work is indicating that a major fraction of the optical and NIR H\,I emission lines are emitted in dense gas ($n$\,$>>$\,10$^{9.5}$\,cm$^{-3}$).

In the RPC models of \citet{baskin14}, both H$\beta$ and C\,III]\,$\lambda$\,1909\r{A} are predominantly produced at radii close to the dust sublimation radius, that is towards the outer part of the BLR (see their Fig. 5). In their canonical model (for which L$_{AGN}$\,=\,10$^{45}$erg/s), C++ is found at column densities through a cloud a little less than neutral hydrogen (see their Fig. 3). For clouds in the outer part of the BLR, the range of columns corresponds to a regime in which the volume density drops rapidly from 10$^{11}$\,cm$^{-3}$ (the density at which the H\,I emission arises) to 10$^9$\,cm$^{-3}$ (see their Fig. 2). Since the critical density for C\,III]\,$\lambda$\,1909\r{A} is $\approx$\,10$^{9.5}$\,cm$^{-3}$, the line emission is collisionally suppressed. Thus the H\,I lines and the C\,III]\,$\lambda$\,1909\r{A} line can arise in the same part of the  BLR, from gas that is at different densities in the same clouds.

The RPC model predicts, without any free parameters, not only that the density of the gas where H\,I lines are produced most strongly is $\sim10^{11}$\,cm$^{-3}$, but that the clouds with suitable conditions are located predominantly halfway to the dust sublimation radius.

\begin{table}
\centering
    \begin{tabular}{l c}
    \hline
    \hline
 Object&  r$_b$\\
      \hline
      MCG--05-14-012& 1.0\\
      MCG--05-23-16 & 0.5\\
      MCG--06-30-015& 1.0\\
      NGC1365 & 0.7\\
      NGC2992 & 0.8\\
      NGC3783 &  0.8\\
      NGC4235 &  1.8\\
      NGC4593 & 0.9\\
      NGC6814 &  0.9\\
      \hline
    \end{tabular}
        \caption{Distance from the nucleus to the broad H\,I emitting region in terms of the dust sublimation radius (see Section~\ref{sec:location} for details).}
    \label{tab:table_radius}
\end{table}

We can test this prediction using the photon fluxes listed in table~\ref{tab:phi}.
These are relatively well constrained to lie in the range 10$^{17.5}$-10$^{18.5}$\,cm$^{-2}$\,s$^{-1}$, comparable to the values found by \cite{korista12} for their four QSOs.
Since we know the 14--195\,keV luminosities for all our targets (list in Table~\ref{table_obj}) we can calculate the distance from the AGN at which the ionising photon flux matches those derived from the line ratios, for a given spectral energy distribution (SED).
The adopted SED is important since it determines the extrapolation of the photon rate from 14\,keV down to 13.6\,eV.
When calculating the photoionisation models, \cite{korista04} used an SED comprising a power law with an additional UV bump as shown in Fig.~1 of \cite{korista97}.
We prefer to adopt the SED derived for NGC\,1068 by \citet{pier94}, since this object is comparable to the AGN we have analysed here.
However, we note that it leads to lower extrapolated photon fluxes, and hence smaller radii, than the SED used by \cite{korista97}.
Finally, in order to scale the results to the dust sublimation radius $R_{dust}$ , we adopt $R_{dust} \sim 0.2$\,pc for an AGN bolometric luminosity of 10$^{45}$\,erg\,s$^{-1}$ as given by \cite{netzer15} for graphite grains.
However, as \cite{netzer15} notes, this distance depends on the size and type of grain as well as the anisotropy of the central radiation source, and therefore $R_{dust}$ should not be considered too strictly as being at a fixed location.

The resulting distances are listed in Table~\ref{tab:table_radius}. With the exception of MCG--05-23-16, for which the photon flux is poorly constrained as it was estimated based on only three line ratios, and NGC\,4235, which have the lowest photon flux in the sample, the distances are consistent with a significant amount of the H\,I emitting gas being located at regions near (or even at) the dust sublimation radius. This matches the expectations of the RPC model, and is also in agreement with other theoretical studies which suggested broad Paschen and Balmer lines in large part come from the outer BLR, near or at the dust sublimation radius \citep{zhu09,czerny11,mor12}.

\subsection{Predicted Line Ratios}

Here we compare the predicted broad line ratios for the objects in our study with measured line ratios in Seyfert 1/QSO samples in previous studies.

\citet{dong08} measured the broad H$\alpha$/H$\beta$ line ratio on a sample of 446 nearby Seyfert 1 and quasi-stellar objects that have blue continua (in order to minimise the effects of dust extinction). The measured broad H$\alpha$/H$\beta$ line ratios range from 2.3 to 4.2, skewed towards the large ratio end. This distribution is well described by a log-Gaussian with a peak of 3.06 and a standard deviation of 0.03\,dex (corrected for measurement errors). It is worth noting that, although the broad H$\alpha$/H$\beta$ ratio varies from 1 to 17 across the parameter plane of gas density and ionising flux characteristic of BLR clouds, LOC integrations over the range of cloud parameters adopted by \citet{korista04} predict H$\alpha$/H$\beta$ in this range (cf. discussions in \citet{dong08}. \citet{lamura07} measured the H$\alpha$/H$\beta$ in 90 nearby (z\,$\le$\,0.368) type 1 AGNs and they obtained a median of 3.45\,$\pm$\,0.65, while the measured ratios ranged from 1.56 to 4.87. \citet{zhou06} measured the broad H$\alpha$/H$\beta$ line ratio in a sample of 2000 narrow-line Seyfert 1s and they obtained an average ratio of H$\alpha$/H$\beta$\,=\,3.03 with a dispersion of $\approx$\,0.36. In a sample of broad line AGN with low starlight contamination, \citet{greene05} found a mean total (narrow and broad) H$\alpha$/H$\beta$ ratio of 3.5.  

The line ratios we observe for the unreddened Seyfert~1s MGC-05--14--012, NGC\,3783 and NGC\,4593 are in good agreement with these other observational studies; as is the predicted line ratio for NGC\,6814, MCG--06--30--015 and MCG-05--23--16, the latter based on the best fit to the H$\alpha$ and Paschen lines, and for which we derived a substantial extinction of $A_V(BLR) = 7.7$\,mag.
In contrast, the dereddened line ratios for NGC\,1365, NGC\,2992, NGC\,4235 are clearly larger than the ratios measured for the objects in the \citet{dong08} and \citet{lamura07} studies.
We point out that although these studies suggest that the median Balmer decrement in the broad line region is around $\approx$\,3, much larger Balmer decrements have been observed in unreddened AGN before, for example, \citet{rudy88} measured a H$\alpha$/H$\beta$ ratio of 5.0 in Mrk\,609 while \citet{osterbrock81} measured a ratio of 7.8 in the same object when it was in a low optical state. 

\citet{lamura07} also studied the H$\gamma$/H$\beta$, obtaining H$\gamma$/H$\beta$\,=\,0.45\,$\pm$\,0.08. \citet{zhu09} analysed a sample of 74 narrow-line Seyfert 1s and found an average H$\gamma$/H$\beta$ ratio of 0.4. The H$\gamma$/H$\beta$ ratios in our objects, whether directly measured, rereddened or predicted from the model fitted to other line ratios, are all in good agreement with these results.

\subsection{Comparison of extinction to previous studies}

In this subsection we compare our extinction estimates -- both to the NLR and the BLR as given in Table~\ref{tab:av} -- with previous estimates available in the literature.
In order to compare published values of E(B-V) with our A$_V$ estimates, we have adopted A$_V$\,=\,3.1\,E(B-V) throughout.
\begin{description}
\item{\textbf{MCG\,--05-23-16}}: In a previous analysis of this galaxy, \citet{veilleux97} estimated A$_V$(NLR)\,=\,0.89 from the H$\alpha$/H$\beta$ and Pa$\beta$/H$\alpha$ line ratios and they estimated a lower limit of A$_V$(BLR)\,=\,7 based on the broad Pa$\beta$ flux and the lack of a broad H$\alpha$ component in their observations, assuming case B recombination line ratios. These results are consistent with ours.
\item{\textbf{MCG\,--06-30-015}}: \citet{reynolds97} estimated A$_V$\,=\,2.0--2.4 from the slope of the nuclear continuum, in agreement with our estimate.
\item{\textbf{NGC\,1365}}: \citet{schulz99} derived an extinction to the broad line region in the range of A$_{V}$(BLR)\,=\,3.5--4 based on the comparison of the observed H$\alpha$ luminosity in the inner few arcsecs and the intrinsic H$\alpha$ luminosity estimated from modelling of narrow emission line ratios. This estimate is consistent with our estimate of A$_{V}$(BLR)\,=\,4.2. \citet{trippe10} obtained long-slit spectra of the inner 2\arcsec\ of NGC\,1365 in January 2009, covering the 3500--7000\r{A} spectra range. They estimated A$_V$\,=\,4.1, much larger than our estimate of A$_{V}$(NLR)\,=\,1.3. To comprehend this difference, we compare their observations to ours: they mention that in their spectrum the narrow H$\beta$ appears to be enhanced relative to the [O\,III] lines, indicating a very strong starburst emission component. This is not observed in our spectrum, which covers only the inner 1\farcs2 of NGC\,1365.
However, within the field of view of the IFU, we do observe an enhancement of the narrow H$\beta$ relative to the [O\,III] lines in a region $\approx$1\farcs8 distant from the nucleus, and therefore beyond our extracted spectrum.
Thus, the difference in estimates of the extinction to the NLR are likely due to the different spatial regions probed by our observations.
\item{\textbf{NGC\,2992}}: \citet{veilleux97} estimated the extinction to the BLR from the Pa$\beta$/Br$\gamma$ line ratio assuming case B recombination and obtained A$_V$\,=\,6.5. \citet{gilli00} estimated the reddening to the broad and narrow line regions assuming case B recombination and found they are both similar, $A_V(NLR)$\,=\,2.0 and A$_V(BLR)$\,=\,2.2. The difference to our results are partly because case B recombination is not a valid assumption in this case. However, \citet{gilli00} also reported observed ratios for H$\alpha$/Pa$\beta$ of 3.3 and for Pa$\beta$(broad)/Pa$\beta$(narrow) of 17, both larger than our measurements of 2.71 and 10.5 respectively. These differences are not consistent with a simple change in obscuration, supporting the study by \cite{trippe08} who argued that the broad line region emission in this galaxy shows variability. \citet{trippe08} observed NGC\,2992 in 2005 and they estimated A$_V$(NLR)\,=\,2.2, close to our estimate of A$_V$(NLR)\,=\,2.8. \citet{trippe08} did not detect broad H$\beta$ or H$\alpha$ lines in their spectra.
\item{\textbf{NGC\,3783}}: \citet{barr83} estimated A$_V$\,=\,0.56$^{+0.37}_{-0.50}$ from the UV continuum, which, like our estimates of $A_V(BLR)$ and $A_V(NLR)$, is consistent with zero extinction.
\item{\textbf{NGC\,4593}}: \citet{ward87b} observed the 0.3--100\,$\mu$m continuum of NGC\,4593 and, based on its shape, concluded that it is a minimally reddened AGN. This matches the zero extinction we find to both the NLR and BLR.
\item{\textbf{NGC\,6814}}: \citet{winkler97} estimated A$_V$\,=\,0.65$\pm0.37$ for the nucleus of NGC\,6814 based on a comparison of the observed B--V,U--B,V--R and V--I colours of the nuclear continuum to those of unreddened AGN. This is consistent with our estimate of A$_V$(BLR)\,=\,0.4$\pm0.4$.
\end{description}

\subsection{Absorption to the BLR and NLR}

\begin{figure}

\includegraphics[scale=0.37,angle=-90]{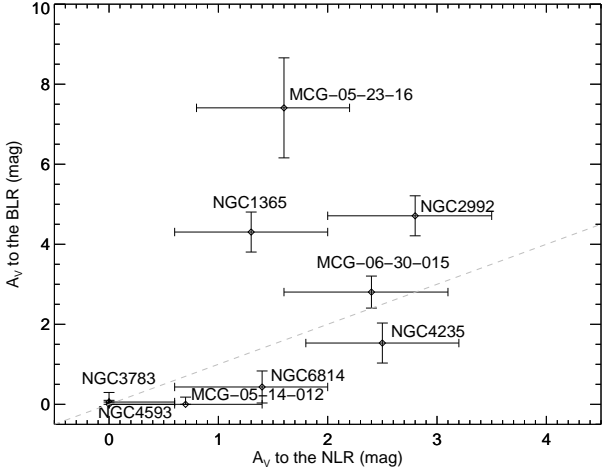}
\caption{Comparison of A$_{V}$ to the broad line region versus A$_{V}$ to the narrow line region. While six objects have similar extinctions to the BLR and to the NLR, three objects have large extinction towards the BLR and moderate extinction to the NLR.}
\label{fig3}
\end{figure}

The torus, which obscures some lines of sight to the AGN, is a central tenet of the AGN unification scheme.
As such, it is usually assumed that all obscuration to an AGN occurs on these very small scales;
and, by extension, that Seyferts suffering intermediate levels of obscuration are viewed through lines of sight that graze the edge of the torus.
Our sample of Seyfert~1s includes several with intermediate obscuration, and so in this section we discuss whether that is necessarily the case.

In Fig.\,\ref{fig3} we compare the extinction to the NLR to the extinction to the BLR. In three sources where the BLR is unobscured, there is also no obscuration to the NLR (within errors), which is expected. Three sources have A$_V$(BLR)\,$\simeq$\,A$_V$(NLR) within errors, which implies in these cases an obscuring structure at larger scales than the NLR. In the Seyfert 1.8/1.9s in our sample, the extinction to the NLR is substantially smaller: 1.9\,mag smaller in NGC\,2992, 2.9\,mag in NGC\,1365 and 6.1\,mag in MGC\,--05-23-16. This implies that the extinction to the BLR is due to dusty structures that are more compact than the NLR. This could be the torus, and, in fact, previous studies of the spectral variation in the broad lines and continuum of Seyfert 1.8/1.9s \citep{goodrich89,tram92,goodrich95} showed that, in some of these objects, the variations were consistent with changes in reddening, and the time scales involved pointed to the obscuring source being close to the BLR, i.e., the torus. On the other hand, dusty nuclear spirals and filaments are also a possibility: in a recent study, \citet{prieto14} found that dusty nuclear spirals and filaments on scales of tens of parsecs to a hundred parsecs can cause moderate extinction to the nucleus, in the range A$_V$\,=\,3--6\,mag.

Such structures can in principle account for the extinction to the BLR in NGC\,1365 and NGC\,2992, but they cannot account for the extinction to the BLR in MCG--05-23-16. This latter object was one of those observed by \citet{prieto14}, who found the filament crossing the nucleus causes only a small amount of extinction, A$_V$\,$\approx$\,1\,mag. This implies the extinction to the BLR of MCG--05-23-16 occurs on a much smaller scale.The mid-IR emission in AGNs tracing the warm dust is typically located on scales $<$\,50\,pc for the AGNs in our luminosity range, and specifically in the case of MCG--05-23-016, the half-light radius of the mid-IR emission is $\approx$2\,pc as found by interferometry \citep{leonard13}. Thus, in the case of MCG--05-23-16, considering the large difference between the extinction to the NLR and BLR, the fact that high spatial resolution observations showed that the dust filament crossing its nucleus causes only a small amount of extinction \citep{prieto14} and the size of the half-light radius of the mid-IR emission \citep{leonard13}, we argue the torus is the structure obscuring the BLR.

In a recent study, \citet{heard16} estimated the reddening to the broad and narrow line regions independently for 300 objects based on the H$\alpha$/H$\beta$ ratio, which they assumed to be equal to 2.9 in the NLR spectra and to 2.78 in the BLR spectra. They find that the BLR is always reddened at least as much as the NLR and often much more (see their Fig.\,3).  In contrast, we find that in six of the nine objects in our sample the extinction to BLR and NLR are the same within errors. We attribute this difference to \citet{heard16} adopting a single intrinsic broad H$\alpha$/H$\beta$ line ratio of 2.78 for all AGN in their analysis, which will result in overestimating the extinction to the BLR in many sources, as our analysis indicates that the intrinsic broad H$\alpha$/H$\beta$ ratio has a broad distribution and is usually larger than this value.

\subsection{X-ray absorption and optical extinction}

\begin{figure}

\includegraphics[scale=0.37,angle=-90]{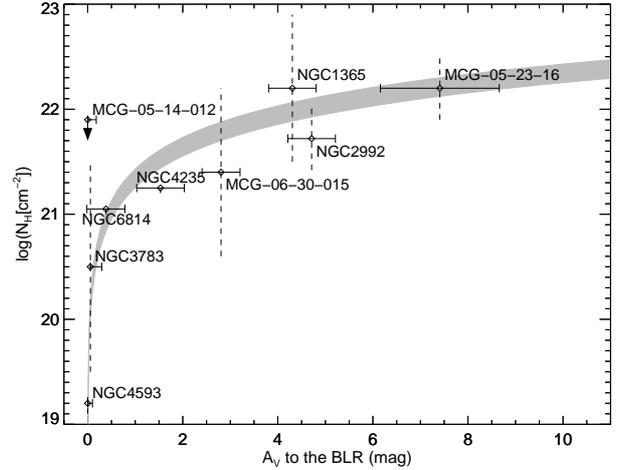}
\caption{A$_{V}$ to the broad line region versus X-ray absorbing column N$_H$, adapted from a similar figure by \citet{leonard15}. The vertical dashed lines denote the range of absorbing columns in the literature, which indicates the minimum variability for each object. The N$_H$ value for MCG\,--05-14-012 is an upper limit. The thick grey line represents the Galactic standard gas-to-dust ratio \citep{predehl95,nowak12}.}
\label{fig4}
\end{figure}

A quantitative comparison between extinction and X-ray absorbing column has recently been presented by \cite{leonard15} -- and we refer to these authors for a discussion of other results in the literature on this topic.
They showed that when one takes into account the variability in the X-ray absorbing column, its lower limit is consistent with the measured optical extinction for the Galactic gas-to-dust ratio; and they argued that deviations from this can be explained as absorption by dust-free neutral gas within the BLR.
The extinction used by these authors was derived using the `dust colour' method of \cite{leonard15a}, and is a measure of the extinction to the hot ($\sim1500$\,K) dust.

Because this probes nearly all the way to the BLR, it can be used as a proxy for the extinction to the BLR. But also because of that, we re-visit this issue here and look at whether their conclusion is upheld when using the extinction derived directly from the broad lines themselves.

We plot A$_V$(BLR) versus the X-ray absorbing columns in Fig.\,\ref{fig4}. The thick grey curve gives the range of relations using the Galactic gas-to-dust ratio (N$_H$/A$_V$)$_{Galactic}$\,=\,(1.79--2.69)\,$\times$10$^{21}$cm$^{-2}$ \citep{predehl95,nowak12}.
The observed range of N$_H$ for each object, tracing variations in X-ray absorbing column, are indicated by dashed lines. When N$_H$ variation is taken into account, all objects in this study show that the lower end of their range of N$_H$/A$_V$ ratios is consistent with the Galactic gas-to-dust ratio.
In particular, we have 6 objects in common with \cite{leonard15}, and the $A_V$ derived here from broad line measurements is consistent with -- and more precise than -- that derived with the dust colour method presented by \cite{leonard15}. As such, our analysis confirms their result.

\subsection{Seyfert sub-type and Extinction to the Broad Line Region}

\begin{figure}

\includegraphics[scale=0.37,angle=90]{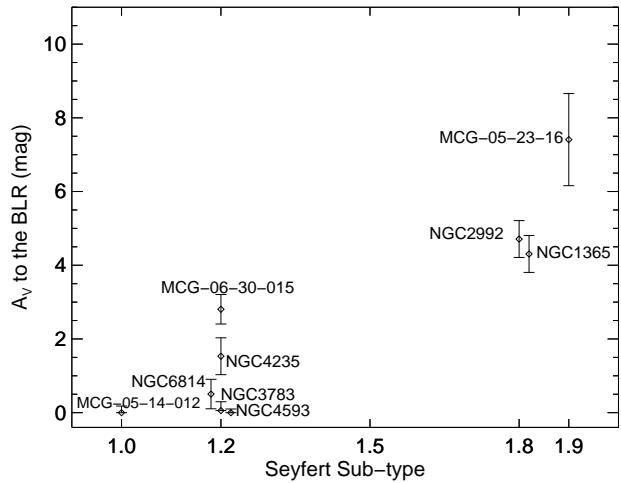}
\caption{Seyfert sub-type versus A$_{V}$ to the broad line region. Sub-types 1.0 are effectively unobscured, Sub-type 1.2 is composed both by unobscured and lightly obscured (0.5--3\,mag) while sub-types 1.8--1.9 suffer substantial obscuration in the range 4--8\,mag. Given Figure~\ref{fig4}, this plot can be considered equivalent to Fig.~2 of \citet{leonard15} which plots sub-type versus X-ray absorbing column.}
\label{fig5}
\end{figure}

Finally, we also look at the relation between Seyfert sub-type (as discussed in Section~\ref{sec:subtypes}) and the extinction to the BLR, which are plotted in Fig.\,\ref{fig5}.
There is a clear separation between Seyfert 1-1.2s and Seyfert 1.8-1.9s: Seyfert 1-1.2s are either unabsorbed objects or mildly absorbed objects, while Seyfert 1.8-1.9s are affected by substantial obscuration to the BLR, in the range 4--8\,mag. Our result is consistent with earlier studies that suggested that these intermediate AGN categories are comprised mostly of partially obscured Seyfert 1s \citep{goodrich89,goodrich95,dong05}, the remaining being objects with low AGN continuum flux \citep{rudy88,trippe10}. 
And it supports the similar comparison by \cite{leonard15} between the Seyfert sub-type and the X-ray absorbing column, which showed a clear separation between the 1.0--1.5 and 1.8/1.9 sub-types.
Indeed, given the relation between X-ray absorbing column and extinction, these two comparisons with the Seyfert sub-type can be considered equivalent.

\section{Summary and Conclusions}
\label{sec:conc}

In this paper, we present simultaneous spectroscopic observations of 6 optical (Balmer) and near-infrared (Paschen) broad lines from nine nearby Seyferts that are part of our ongoing survey of local luminous AGN with matched analogs (LLAMA).
Analysing the H\,I line ratios in the context of a grid of photoionisation models (and without assuming case B recombination), we are able to derive the intrinsic line ratios, photon flux and density of the BLR, as well as the extinction to it. Based on the analysis of these objects, our main conclusions are:

\begin{itemize}
\item Values of the intrinsic H$\alpha$/H$\beta$ line ratio lie in the range 2.5--6.6 (without accounting for uncertainties), confirming that case B recombination cannot be assumed, and that a single set of excitation properties cannot be used to represent all BLRs.
\item The density of the H\,I emitting gas in the BLR is typically 10$^{11}$\,cm$^{-3}$ for the sample, although there may be variation between individual sources. The ionising photon flux is in the range $\sim10^{17.5}$--10$^{18.5}$\,cm$^{-2}$\,s$^{-1}$, which implies that a significant amount of the H\,I emitting gas is located at regions near the dust sublimation radius.
These values are consistent with theoretical predictions, in particular with the radiation pressure confinement model.
\item The Seyfert sub-types, determined via the [O\,III]/H$\beta$ line ratio, are consistent with the extinction to the BLR, which is based on independent estimates from the H\,I and He\,II lines: in Seyfert 1.0 and 1.2s the BLR is either unobscured or mildly obscured while in Seyfert 1.8 and 1.9s the BLR is moderately obscured with extinction in the range A$_V = 4$--8\,mag. By inference, Seyfert~2s have A$_V > 8$\,mag.
\item In the moderately obscured objects the extinction to the BLR is significantly larger than the extinction to the NLR. This may be caused by the torus but could also be due to dusty filaments or a nuclear spiral. In the specific case of MCG--05-23-16, the nuclear filament cannot account for the obscuration, which is instead likely due to the torus.
\end{itemize}

\section*{ACKNOWLEDGEMENTS}

This research has made use of the SIMBAD database, operated at CDS, Strasbourg, France and of the NASA/IPAC Extragalactic Database (NED) which is operated by the Jet Propulsion Laboratory, California Institute of Technology, under contract with the National Aeronautics and Space Administration. This research has also made use of NASA’s Astrophysics Data System Bibliographic Services.
ASM is supported by Brazilian institution CAPES. LB is supported by a DFG grant within the SPP “Physics of the interstellar medium”. 

\bibliographystyle{mnras.bst}
\bibliography{paper.bib}

\begin{figure*}

\includegraphics[scale=0.5,angle=-90]{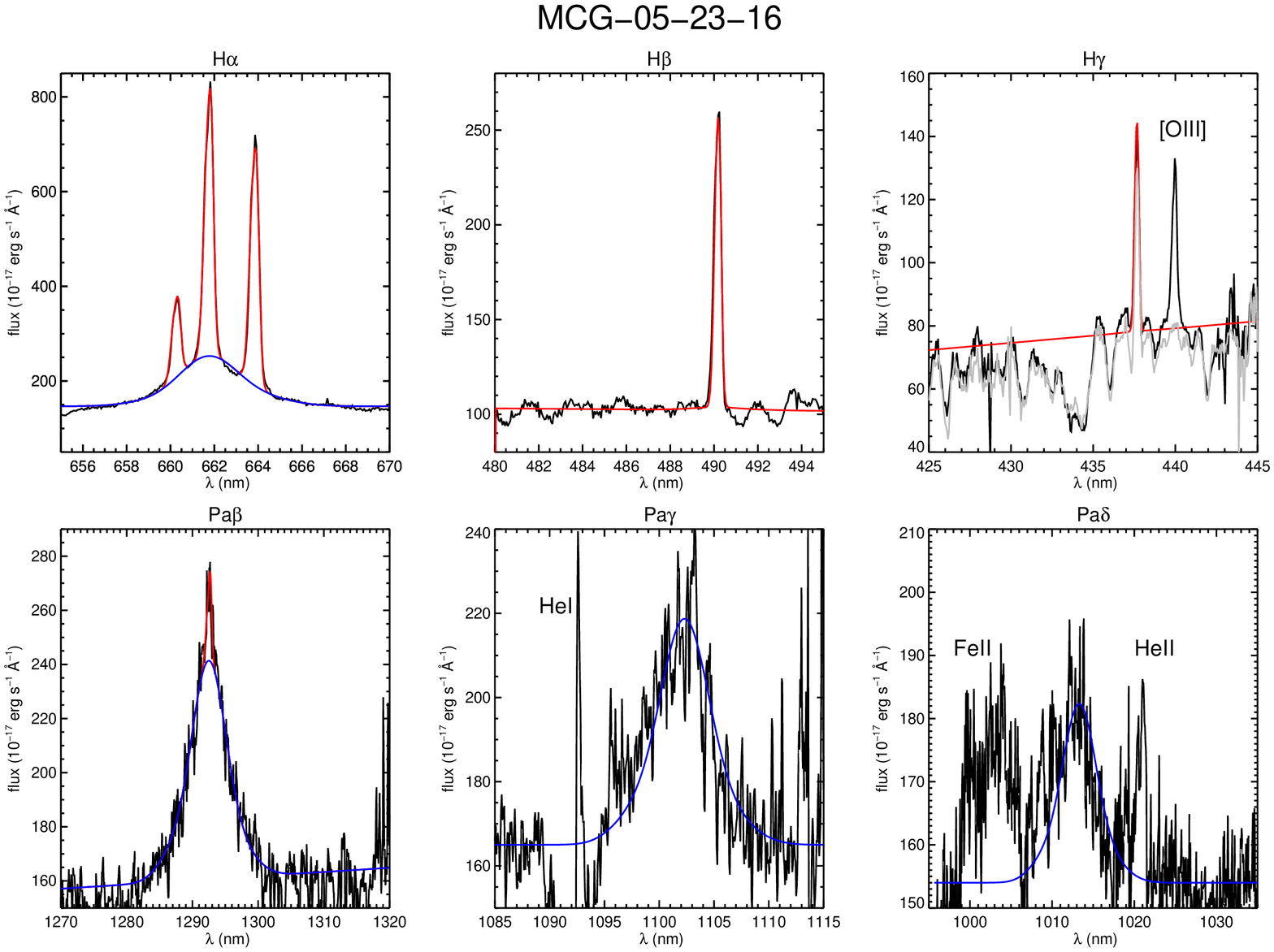}
\caption{Observed H$\alpha$, H$\beta$, H$\gamma$, Pa$\beta$, Pa$\gamma$ and Pa$\delta$ line profiles from an integrated spectrum of the inner 1\farcs2 (black), fitted broad (blue) and broad plus narrow (red) profiles lines. On the H$\gamma$ panel, we show in grey a combination of the fitted H$\gamma$ profile and the spectrum of an inactive galaxy, in order to illustrate that most of the differences between the observed and fitted profiles are due to the shape of the underlying stellar continuum. On the Pa$\gamma$ panel, broad He\,I was subtracted from the spectrum.}
\label{fig:mcg0523}
\end{figure*}

\begin{figure*}

\includegraphics[scale=0.5,angle=-90]{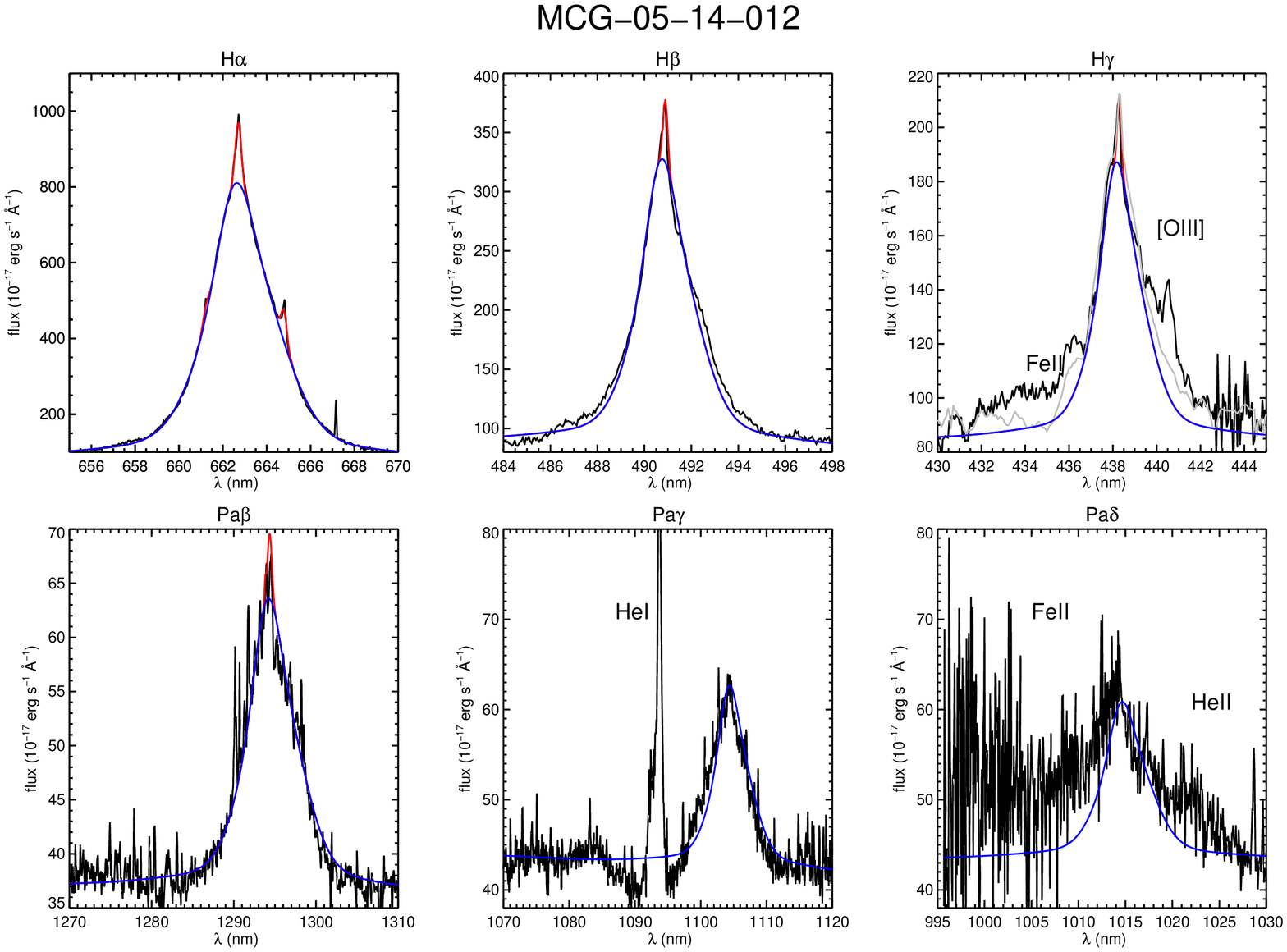}
\caption{Observed H$\alpha$, H$\beta$, H$\gamma$, Pa$\beta$, Pa$\gamma$ and Pa$\delta$ line profiles from an integrated spectrum of the inner 1\farcs2  (black), fitted broad (blue) and broad plus narrow (red) profiles lines. On the H$\gamma$ panel, we show in grey a combination of the fitted H$\gamma$ profile and the spectrum of an inactive galaxy, in order to illustrate that most of the differences between the observed and fitted profiles are due to the shape of the underlying stellar continuum. Excess emission on the wings of the broad H$\gamma$ line are due to [O\,III] and Fe\,II multiplets. On the Pa$\gamma$ panel, broad He\,I was subtracted from the spectrum.}
\label{fig:mcg0514}
\end{figure*}

\begin{figure*}

\includegraphics[scale=0.5,angle=-90]{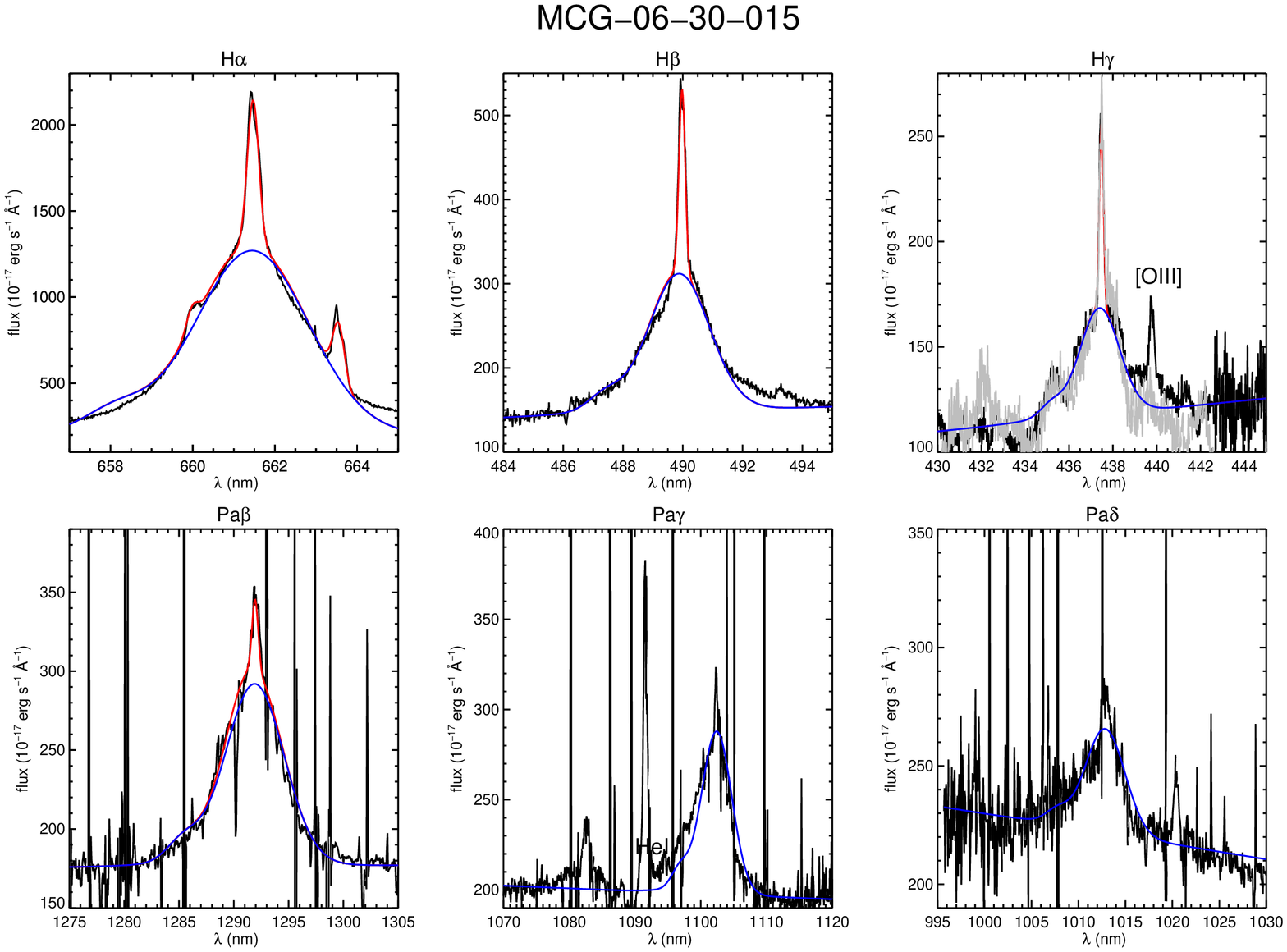}
\caption{Observed H$\alpha$, H$\beta$, H$\gamma$, Pa$\beta$, Pa$\gamma$ and Pa$\delta$ line profiles from an integrated spectrum of the inner 1\farcs2  (black), fitted broad (blue) and broad plus narrow (red) profiles lines. On the H$\gamma$ panel, we show in grey a combination of the fitted H$\gamma$ profile and the spectrum of an inactive galaxy, in order to illustrate that most of the differences between the observed and fitted profiles are due to the shape of the underlying stellar continuum. On the Pa$\gamma$ panel, broad He\,I was subtracted from the spectrum.}
\label{fig:mcg0630}
\end{figure*}

\begin{figure*}

\includegraphics[scale=0.5,angle=-90]{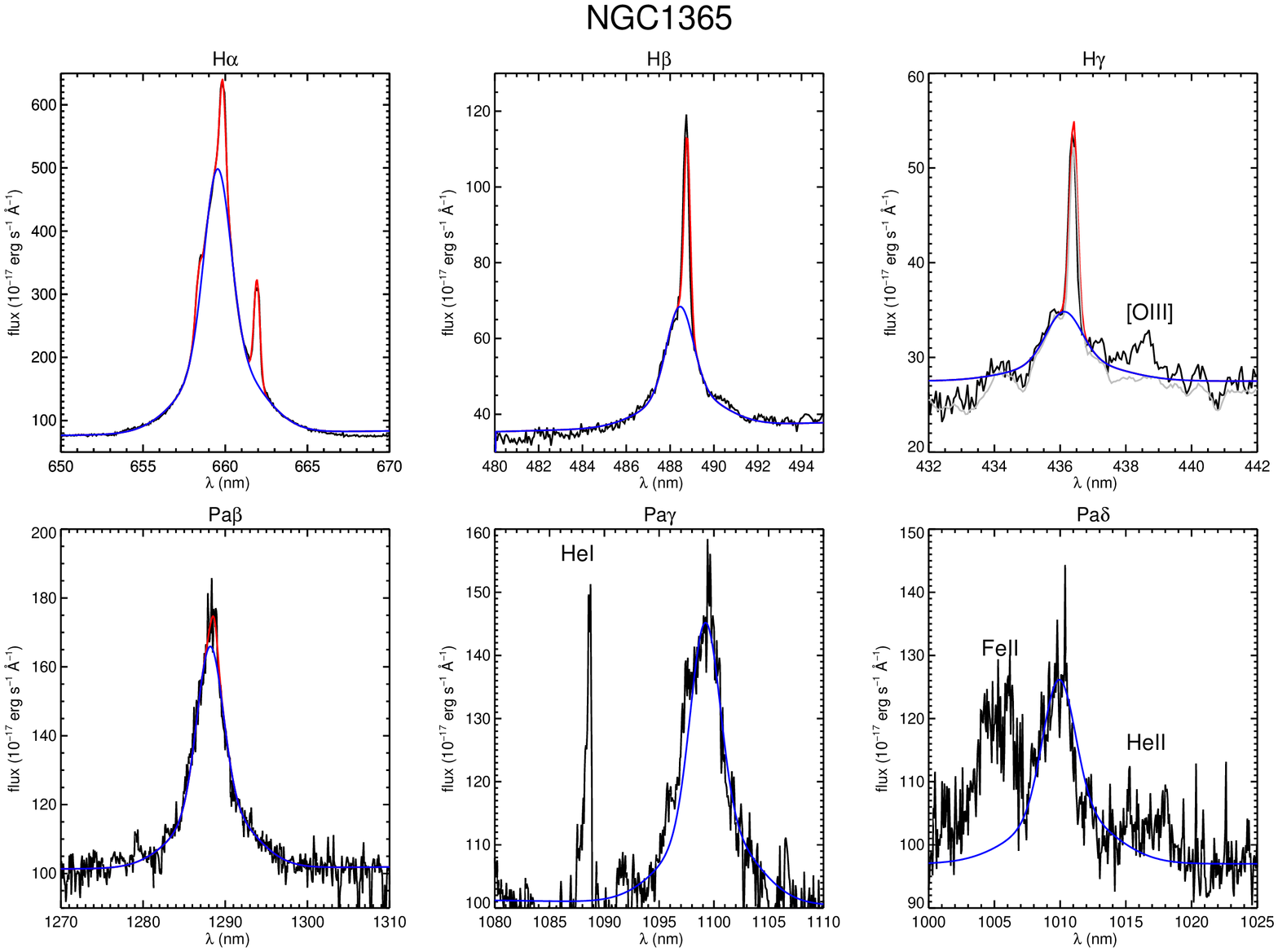}
\caption{Observed H$\alpha$, H$\beta$, H$\gamma$, Pa$\beta$, Pa$\gamma$ and Pa$\delta$ line profiles from an integrated spectrum of the inner 1\farcs2  (black), fitted broad (blue) and broad plus narrow (red) profiles lines. On the H$\gamma$ panel, we show in grey a combination of the fitted H$\gamma$ profile and the spectrum of an inactive galaxy, in order to illustrate that most of the differences between the observed and fitted profiles are due to the shape of the underlying stellar continuum. On the Pa$\gamma$ panel, broad He\,I was subtracted from the spectrum.}
\label{fig:ngc1365}
\end{figure*}

\begin{figure*}

\includegraphics[scale=0.5,angle=-90]{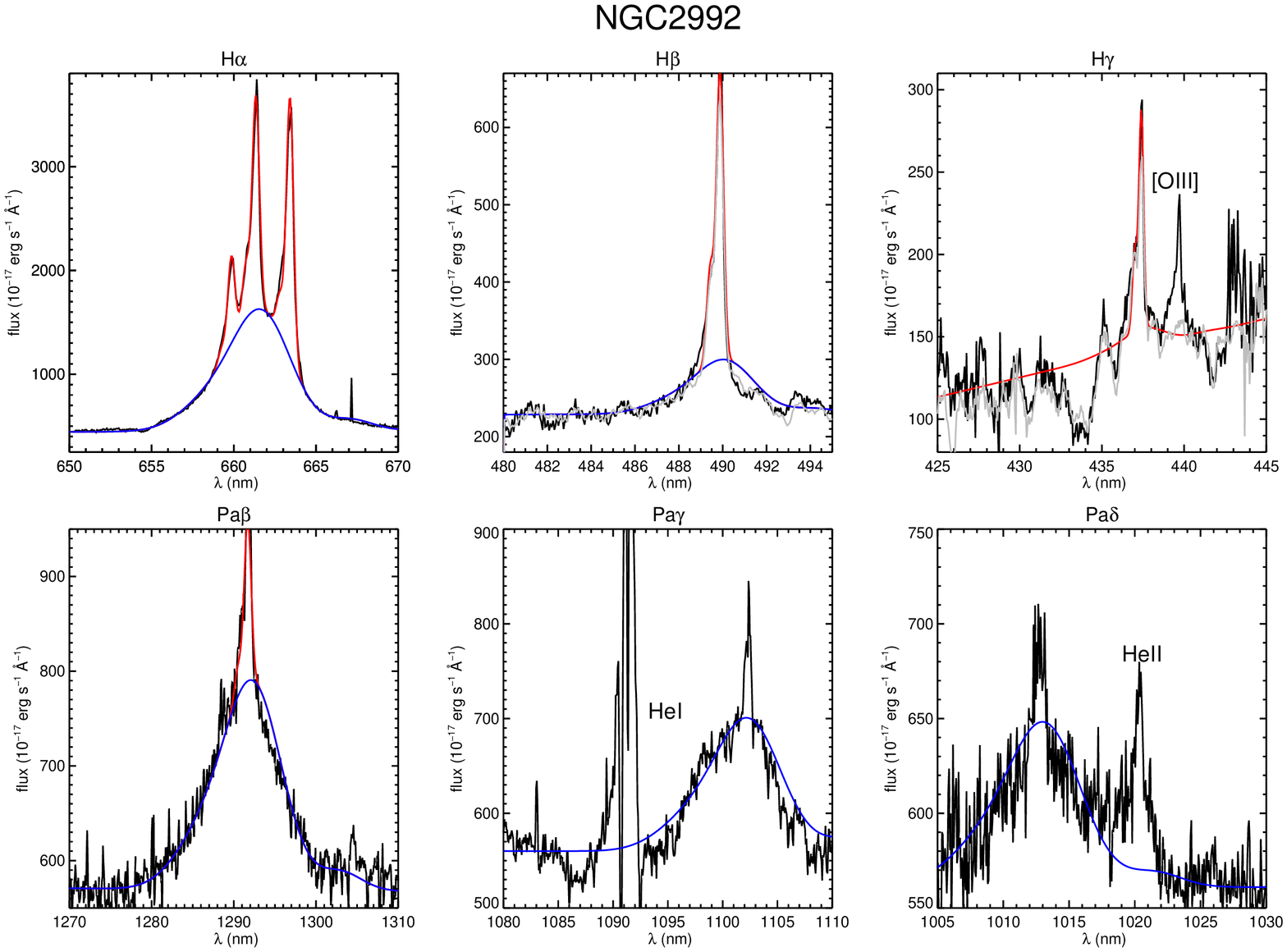}
\caption{Observed H$\alpha$, H$\beta$, H$\gamma$, Pa$\beta$, Pa$\gamma$ and Pa$\delta$ line profiles from an integrated spectrum of the inner 1\farcs2  (black), fitted broad (blue) and broad plus narrow (red) profiles lines. On the H$\gamma$ panel, we show in grey a combination of the fitted H$\gamma$ profile and the spectrum of an inactive galaxy, in order to illustrate that most of the differences between the observed and fitted profiles are due to the shape of the underlying stellar continuum. On the Pa$\gamma$ panel, broad He\,I was subtracted from the spectrum.}
\label{fig:ngc2992}
\end{figure*}

\begin{figure*}

\includegraphics[scale=0.5,angle=-90]{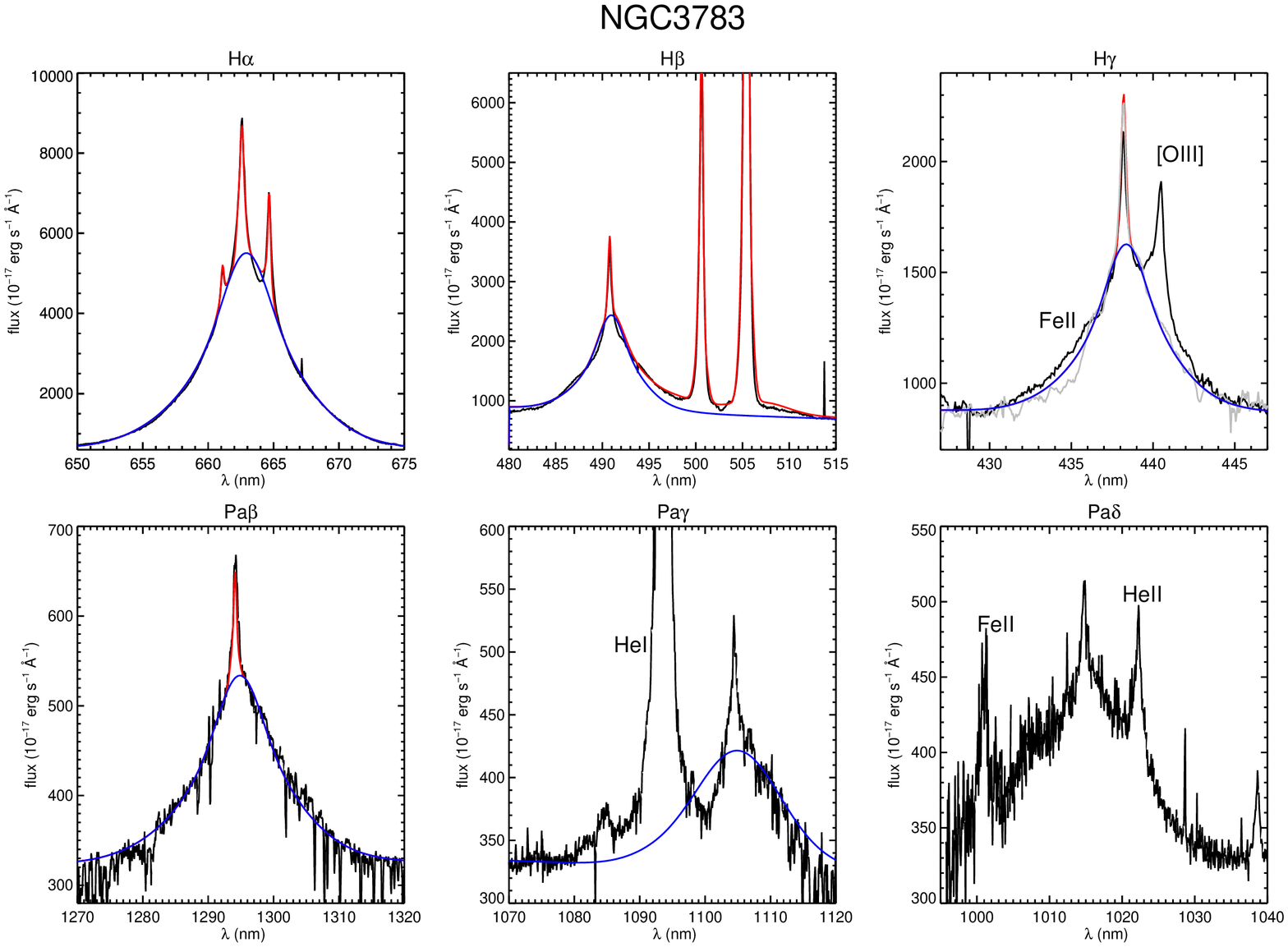}
\caption{Observed H$\alpha$, H$\beta$, H$\gamma$, Pa$\beta$, Pa$\gamma$ and Pa$\delta$ line profiles from an integrated spectrum of the inner 1\farcs2  (black), fitted broad (blue) and broad plus narrow (red) profiles lines. On the H$\gamma$ panel, we show in grey a combination of the fitted H$\gamma$ profile and the spectrum of an inactive galaxy, in order to illustrate that most of the differences between the observed and fitted profiles are due to the shape of the underlying stellar continuum. Excess emission on the wings of the broad H$\gamma$ line are due to [O\,III] and Fe\,II multiplets. On the Pa$\gamma$ panel, broad He\,I was subtracted from the spectrum. Broad Pa$\delta$ was not fitted as the line could not be deblended from from He\,II\,$\lambda$10126.4\r{A} and Fe\,II\,$\lambda\lambda$9956,9998\r{A}.}
\label{fig:ngc3783}
\end{figure*}

\begin{figure*}

\includegraphics[scale=0.5,angle=-90]{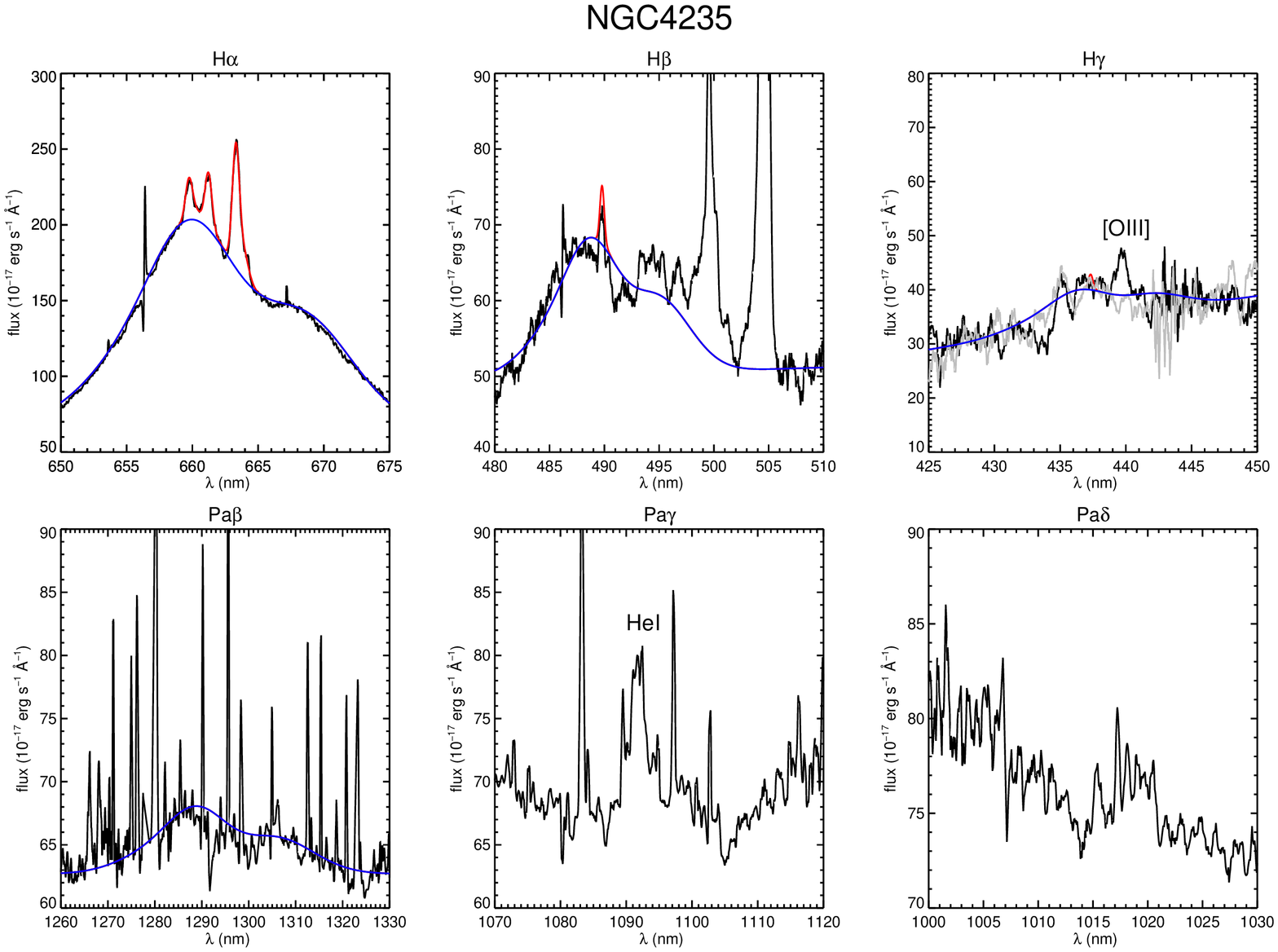}
\caption{Observed H$\alpha$, H$\beta$, H$\gamma$, Pa$\beta$, Pa$\gamma$ and Pa$\delta$ line profiles from an integrated spectrum of the inner 1\farcs2  (black), fitted broad (blue) and broad plus narrow (red) profiles lines. On the H$\gamma$ panel, we show in grey a combination of the fitted H$\gamma$ profile and the spectrum of an inactive galaxy, in order to illustrate that most of the differences between the observed and fitted profiles are due to the shape of the underlying stellar continuum. On the Pa$\gamma$ panel, broad He\,I was subtracted from the spectrum. Broad Pa$\delta$, Pa$\gamma$ and narrow Pa$\beta$ were not detected in this object.}
\label{fig:ngc4235}
\end{figure*}

\begin{figure*}

\includegraphics[scale=0.5,angle=-90]{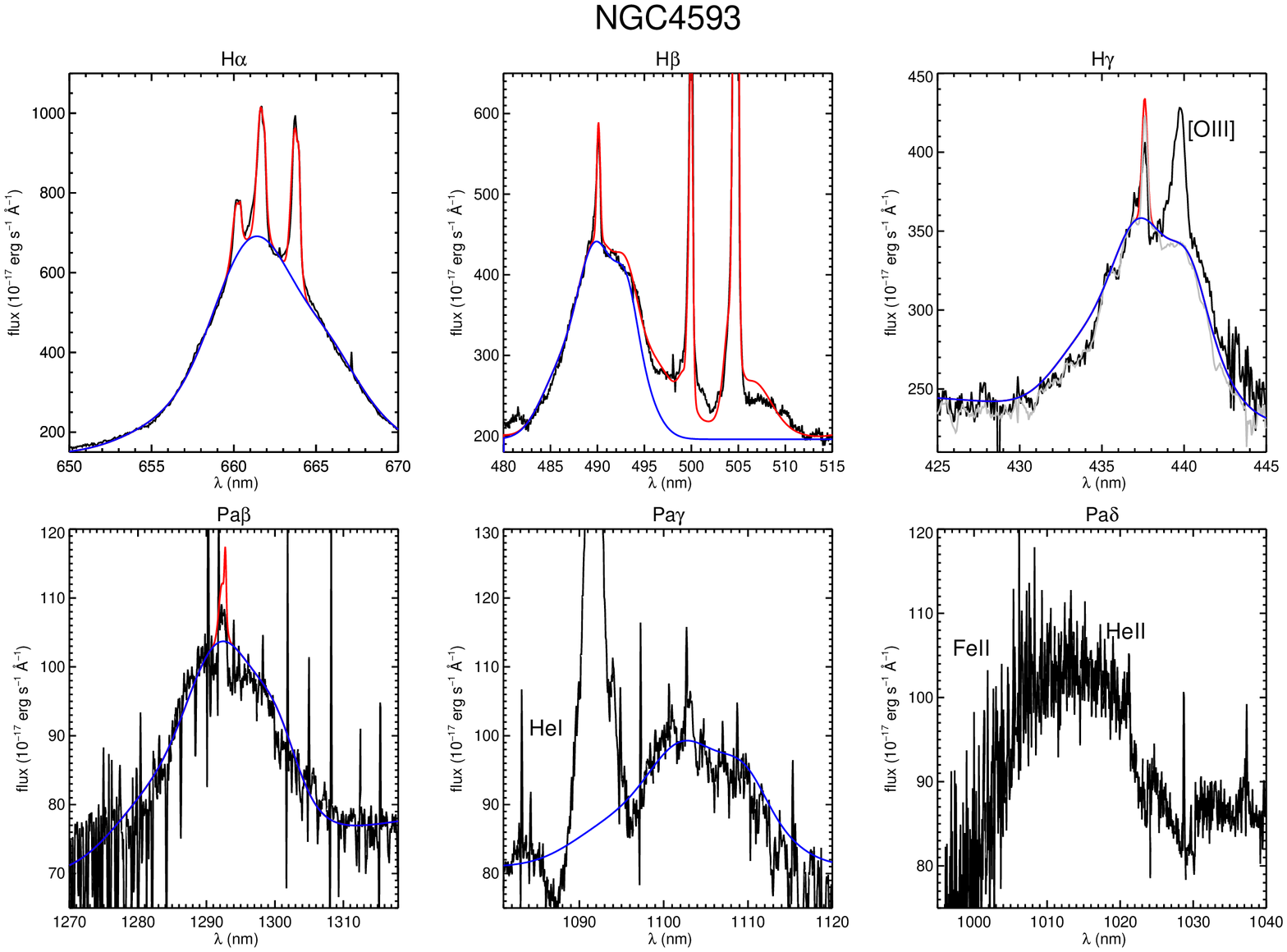}
\caption{Observed H$\alpha$, H$\beta$, H$\gamma$, Pa$\beta$, Pa$\gamma$ and Pa$\delta$ line profiles from an integrated spectrum of the inner 1\farcs2  (black), fitted broad (blue) and broad plus narrow (red) profiles lines. On the H$\gamma$ panel, we show in grey a combination of the fitted H$\gamma$ profile and the spectrum of an inactive galaxy, in order to illustrate that most of the differences between the observed and fitted profiles are due to the shape of the underlying stellar continuum. On the Pa$\gamma$ panel, broad He\,I was subtracted from the spectrum. Broad Pa$\delta$ was not fitted as the line could not be deblended from from He\,II\,$\lambda$10126.4\r{A} and Fe\,II\,$\lambda\lambda$9956,9998\r{A}.}
\label{fig:ngc4593}
\end{figure*}

\begin{figure*}

\includegraphics[scale=0.5,angle=-90]{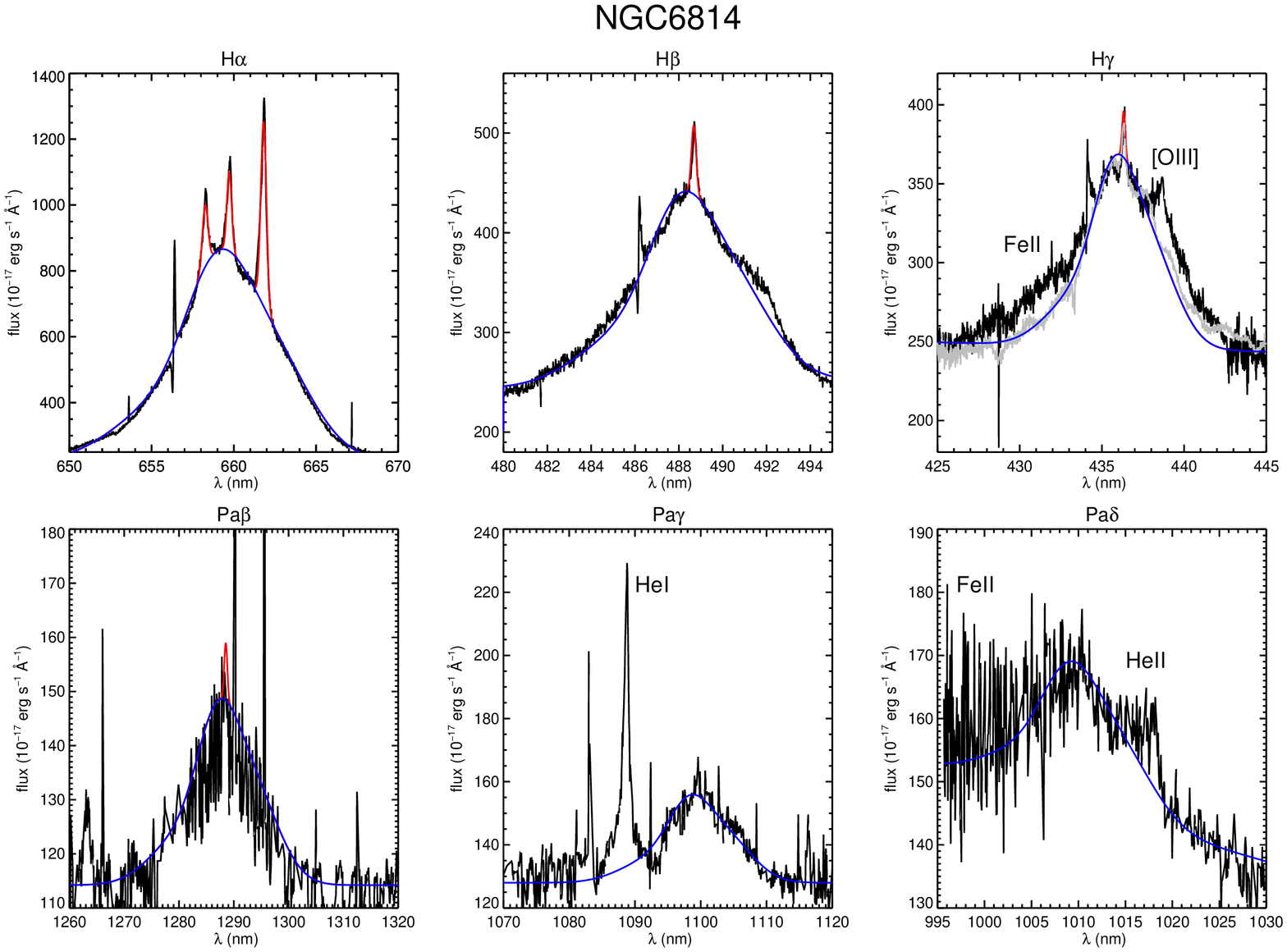}
\caption{Observed H$\alpha$, H$\beta$, H$\gamma$, Pa$\beta$, Pa$\gamma$ and Pa$\delta$ line profiles from an integrated spectrum of the inner 1\farcs2  (black), fitted broad (blue) and broad plus narrow (red) profiles lines.  On the H$\gamma$ panel, we show in grey a combination of the fitted H$\gamma$ profile and the spectrum of an inactive galaxy, in order to illustrate that most of the differences between the observed and fitted profiles are due to the shape of the underlying stellar continuum. Excess emission on the wings of the broad H$\gamma$ line are due to [O\,III] and Fe\,II multiplets. On the Pa$\gamma$ panel, broad He\,I was subtracted from the spectrum.}
\label{fig:ngc6814}
\end{figure*}

\label{lastpage}
\end{document}